\documentclass[twocolumn,twocolappendix]{aastex63}
\usepackage{graphicx}
\usepackage{upgreek}

\newcommand{\oii}{[O\,II] }
\newcommand{\oiirm}{\mathrm{[O\,II]}}
\newcommand{\oiii}{[O\,III] }
\newcommand{\nuv}{\mathrm{NUV}}

\usepackage{amsmath}
\DeclareMathOperator\erf{erf}
\DeclareMathOperator\sgn{sgn}

\received{XXX}
\revised{YYY}
\accepted{ZZZ}

\submitjournal{ApJ}

\shorttitle{Distribution Function and Luminosity Function of \oii}
\shortauthors{Gao \& Jing}

\begin{document}

\title{Universal conditional distribution function of [O\,II] luminosity of galaxies, and prediction for the [O\,II] luminosity function at redshift $z<3$}

\correspondingauthor{Y.P. Jing}
\email{ypjing@sjtu.edu.cn}

\author{Hongyu Gao}
\affiliation{Department of Astronomy, School of Physics and Astronomy, Shanghai Jiao Tong University, Shanghai 200240, People's Republic of China}
\author[0000-0002-4534-3125]{Y.P. Jing}
\affiliation{Department of Astronomy, School of Physics and Astronomy, Shanghai Jiao Tong University, Shanghai 200240, People's Republic of China}
\affiliation{Tsung-Dao Lee Institute, and Shanghai Key Laboratory for Particle Physics and Cosmology, Shanghai Jiao Tong University, Shanghai 200240, People's Republic of China}

\begin{abstract}
The star-forming emission line galaxies (ELGs) with strong [O\,II] doublet are one of the main spectroscopic targets for the ongoing and upcoming fourth generation galaxy redshift surveys. In this work, we measure the [O\,II] luminosity $L_{\mathrm{[O\,II]}}$ and the near-ultraviolet band absolute magnitude $M_{\mathrm{NUV}}$ for a large sample of galaxies in the redshift range of $0.6\leq z <1.45$ from the Public Data Release 2 (PDR-2) of the VIMOS Public Extragalactic Redshift Survey (VIPERS). We aim to construct the intrinsic relationship between the $L_{\mathrm{[O\,II]}}$ and $M_{\mathrm{NUV}}$ through Bayesian analysis. In particular, we develop two different methods to properly correct for the incompleteness effect and observational errors in the [O\,II] emission line measurement. Our results indicate that the conditional distribution of $L_{\mathrm{[O\,II]}}$ at a given $M_{\mathrm{NUV}}$ can be well described by a universal probability distribution function (PDF), which is independent of $M_{\mathrm{NUV}}$ or redshift. Convolving the $L_{\mathrm{[O\,II]}}$ conditional PDF with the NUV Luminosity function (LF) available in the literature, we make a prediction for [O\,II] LFs at $z<3$. The predicted [O\,II] LFs are in good agreement with the observational results from the literature. Finally, we utilize the predicted [O\,II] LFs to estimate the number counts of [O\,II] emitters for the Subaru Prime Focus Spectrograph (PFS) survey. This universal conditional PDF of $L_{\mathrm{[O\,II]}}$ provides a novel way to optimize the source targeting strategy for [O\,II] emitters in future galaxy redshift surveys, and to model  [O\,II] emitters in theories of galaxy formation.
\end{abstract}


\section{Introduction} \label{sec:intro}
Understanding the accelerating expansion of the Universe is a central challenge for modern cosmology. In the current standard cosmological model, dark energy, which is introduced as a new form of energy, is thought to be responsible for the acceleration of the cosmic expansion. As powerful cosmological probes, the baryon acoustic oscillation \citep[BAO, e.g.,][]{2005MNRAS.362..505C,2005ApJ...633..560E} and the redshift space distortion \citep[RSD, e.g.,][]{1987MNRAS.227....1K} are expected to put tight constraints on the dark energy model as well as to test gravity theories through measuring the Hubble expansion rate and the structure growth rate respectively. Both probes require a sufficient number of galaxies with redshift in a large cosmic volume in order to minimize statistical errors. Making such a redshift survey is the goal of ongoing Stage-\uppercase\expandafter{\romannumeral4} projects such as the Dark Energy Spectroscopic Instrument \citep[DESI,][]{2016arXiv161100036D}, the Subaru Prime Focus Spectrograph \citep[PFS,][]{2014PASJ...66R...1T}, 4-metre Multi-Object Spectroscopic Telescope \citep[4MOST,][]{2019Msngr.175....3D}, Euclid \citep{2011arXiv1110.3193L} and Wide Field Infrared Survey Telescope \citep[WFIRST,][]{2012arXiv1208.4012G,2015arXiv150303757S}. 

For the purpose of studying the dark energy at medium and high redshifts, emission line galaxies (ELGs) whose spectra show significant emission line features are chosen as the primary targets for the redshift surveys. Among the emission lines commonly seen in galaxy spectra such as H$\upalpha$ $(\lambda 6563)$, \oiii $(\lambda 4959, \lambda 5006)$, H$\upbeta$ $(\lambda 4861)$ and \oii $(\lambda 3726, \lambda 3729)$, the forbidden \oii line is one of the most prominent spectral lines because of its pronounced doublet shape, its strength,  and its blue location in the rest-frame. For instance, DESI \citep{2016arXiv161100036D} plans to target a large number of ELGs with strong \oii flux in $0.6<z<1.6$ to achieve a high surface density of 2400 $\mathrm{deg^{-2}}$, while PFS \citep{2014PASJ...66R...1T} aims to observe \oii emitters in $0.8<z<2.4$ spanning a comoving volume $\sim 9\left(\mathrm{Gpc}/h\right)^3$. Before the spectroscopic observation, the \oii emitter candidates will be pre-selected based on their photometric properties. In addition, the \oii emission line is also an important indicator for the star formation
rate (SFR) especially at high redshift \citep[e.g.,][]{1998ARA&A..36..189K, 2004AJ....127.2002K, 2006ApJS..164...81M}. Therefore, the \oii luminosity function (LF), which describes the volume number density of \oii emitters in a given luminosity bin, plays a significant role in effectively planning the future ELGs surveys and studying the theory of galaxy formation. In the last two decades, through both the spectroscopic observations and narrow-band imaging, the \oii LF has been measured at different redshift \citep[e.g.,][]{2002ApJ...570L...1G,2003A&A...402...65H,2003ApJ...589..704T,2005A&A...440...61R,2007ApJ...657..738L,2007ApJS..172..456T,2009A&A...495..759A,2009ApJ...701...86Z,2010MNRAS.405.2594G,2011MNRAS.413.2883B,2012MNRAS.420.1926S,2013MNRAS.428.1128S,2015MNRAS.451.2303S,2013ApJ...769...83C,2013MNRAS.433..796D,2015A&A...575A..40C,2016MNRAS.461.1076C,2015MNRAS.452.3948K,2018PASJ...70S..17H,2020MNRAS.494..199S}, though the determination remains very uncertain at redshift $z>1.5$.

Motivated by making a precise forecast for the expected number density of \oii emitters, predicting the \oii LF as well as its redshift evolution is the main goal of this study. In the last five years, a series of observational studies have been carried out to measure the LF not only for  the \oii line but also for other emission lines. For instance, \cite{2015ApJ...811..141M} develop the H$\upalpha$-\oiii bivariate line luminosity function for the ELGs data of the
WFC3 Infrared Spectroscopic Parallel Survey \citep[WISP,][]{2010ApJ...723..104A, 2011ApJ...743..121A}, and predict the H$\upalpha$ LF at $z\sim 2$ using \oiii emitters. Also for the H$\upalpha$ emission line, \cite{2016A&A...590A...3P} constructed three empirical models that can be used to estimate H$\upalpha$ LF at given redshift. Using the multi-color photometry sample, \cite{2017MNRAS.472.4878V} predict the H$\upalpha$, H$\upbeta$, \oii and \oiii line flux based on their SFR, stellar mass and some empirical recipes in order to compute the number counts for these ELGs. \cite{2019MNRAS.489.2355D} connect the \oiii + H$\upbeta$ flux with UV luminosity, and use this relation to predict \oiii + H$\upbeta$ LF at $z\sim 8$. Recently, \cite{2020MNRAS.494..199S} model the emission-line flux for H$\upalpha$ and \oii by extracting information from the galaxy spectral energy distribution (SED) and then predict their number counts in the up-coming WFIRST and PFS surveys. Moreover, in addition to these observational studies, simulation and semi-analytic models (SAM) are also used to predict the emission line LFs \citep[e.g.,][]{2015MNRAS.454..269P, 2018MNRAS.474..177M, 2019MNRAS.490.3667Z, 2020MNRAS.497.5432F}. Particularly, \cite{2020MNRAS.497.5432F} investigate the linear scaling relations between \oii luminosity and other global galactic properties including SFR, age, stellar mass, u and g band magnitude and use these proxies to estimate the \oii LF.

In view of the fact that young, massive stars tend to produce intense UV radiation that can photo-ionizes neutral oxygen atoms in the ionized regions \citep[e.g.,][]{oesterbrock1974astrophysics,2011piim.book.....D}, the \oii emission line should be directly and tightly related with UV radiation. Accordingly, in this study, we attempt to construct the intrinsic relationship between the \oii luminosity $L_{\oiirm}$ and near-ultraviolet (NUV) band absolute magnitude $M_{\nuv}$ using a large number of emission line galaxies from the VIMOS Public Extragalactic Redshift Survey \citep[VIPERS\footnote{http://vipers.inaf.it},][]{2014A&A...566A.108G,2014A&A...562A..23G,2018A&A...609A..84S}. Compared with the previous studies, we have paid our special attention to the incompleteness of faint \oii emitters in the observation. The incompleteness changes with the intrinsic line strength and with the redshift.  We develop a statistical model to characterize this relation and derive its parameters from our sample with two different methods that properly correct the incompleteness effect. We find that after the incompleteness is properly corrected for, the intrinsic relation between the \oii line and the NUV rest-frame magnitude is universal for galaxies across the redshift between 0.6 and 1.1. With this universal relation, we  predict the \oii LFs from NUV LFs at redshift $z<3$, and find that our predicted \oii LFs are broadly in good agreement with observed \oii LFs in the literature, though the observed ones still have large uncertainties. The intrinsic relation will also be very useful for theoretically understanding the formation of \oii lines in galaxies.

This paper is arranged as follows. We first introduce the observational data set and the \oii flux measurement in Section \ref{sec:data}. Then we illustrate the model and fitting approach in Section \ref{sec:methodology}. The main results and the prediction of \oii LF are presented in Section \ref{sec:results}. Finally, we make a summary in Section \ref{sec:summary}. The cosmological parameters assumed throughout the paper are $\Omega_{\Lambda,0} = 0.7$, $\Omega_{m,0} = 0.3$ and $H_0 = 70\,\mathrm{km s^{-1} Mpc^{-1}}$ at $z=0$.

\begin{figure*}
	\centering
	\includegraphics[scale=1.0]{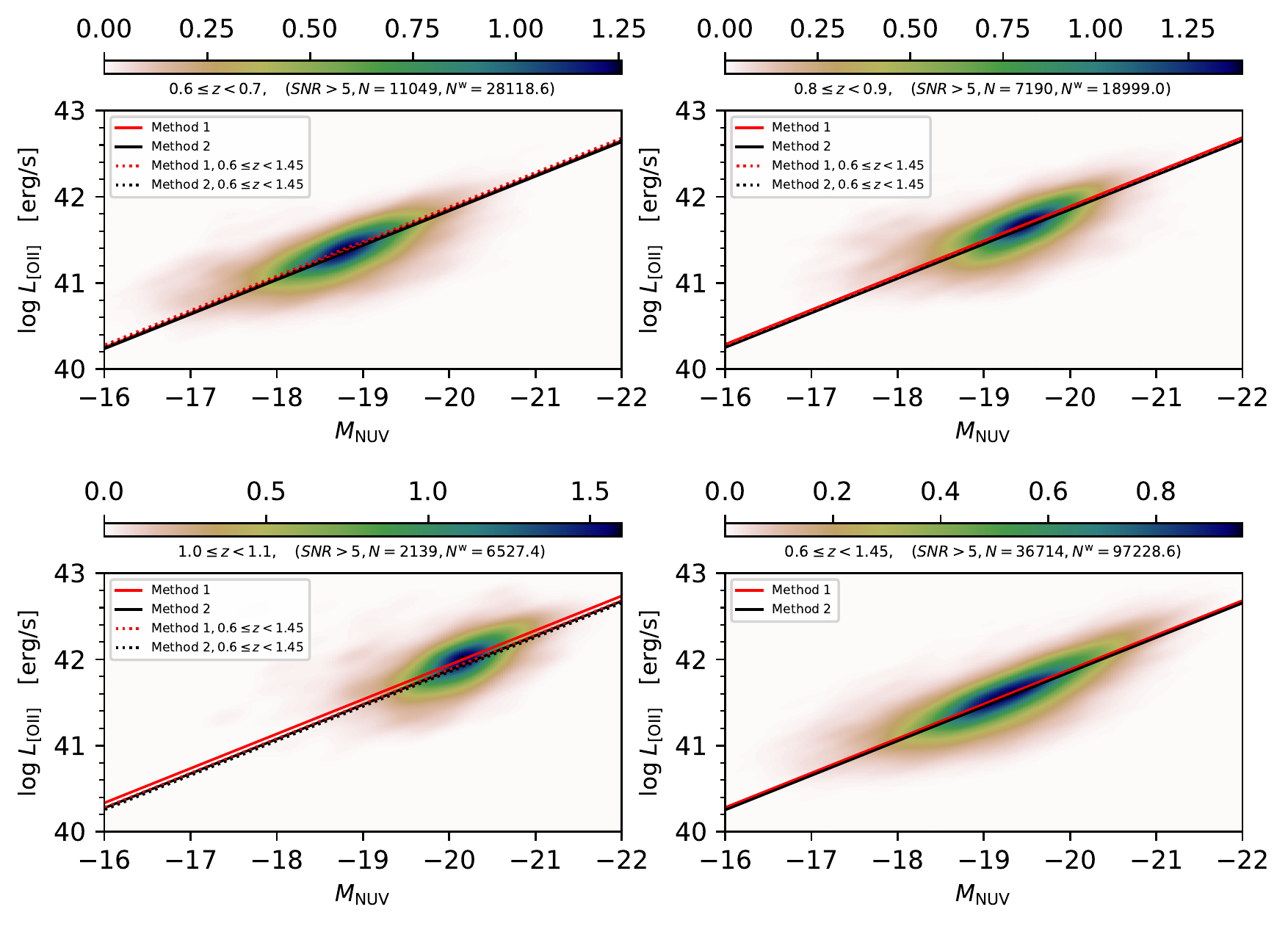}
	\caption{Two-dimensional joint probability distribution function (PDF) of $\log L_{\oiirm}$ and $M_{\nuv}$. Only galaxies with the $L_{\oiirm}$ signal to noise ratio ($SNR$) greater than five are included. The color bar shows the value of the PDF at each pixel, which is calculated based on the kernel density estimate (KDE) with a Gaussian kernel. The first three panels display the PDF in different redshift bins and the last one presents that in the range of $0.6 \leq z<1.45$. The number $N$ and weighted number $N^w$ of the galaxies used are also shown in the title of each panel. The red (method 1) and black (method 2) lines, which often overlap each other, represent the expectation values $\log L^{\mathrm{exp}}_{\oiirm}$ at which the PDF of $L_{\oiirm}$ reaches the maximum at given $M_{\nuv}$ in the model (cf. Section \ref{subsec:Intrinsic}.). The $\log L^{\mathrm{exp}}_{\oiirm}$ is numerically calculated via Equation \ref{equ:xi} and Equation \ref{equ:m0} with our best-fit parameters shown in Table \ref{tab:result}. Meanwhile, the $\log L^{\mathrm{exp}}_{\oiirm}$ of the last panel is also plotted as the dotted lines in the first three panels, which show little evolution of the $L_{\oiirm}$-$M_{\nuv}$ intrinsic relation over the redshift range probed. 
\label{fig:L_M_obs_2d}}
\end{figure*}

\section{Data} \label{sec:data}
In this section, we briefly introduce the galaxy samples used in this study. The \oii emission line fluxes and near-ultraviolet band absolute magnitudes $M_{\nuv}$ of these galaxies are measured through analyzing the spectroscopic data and multi-band photometric data, respectively.

\subsection{VIPERS} \label{subsec:VIPERS}
The galaxy sample is selected from the second data release \citep{2018A&A...609A..84S} of the VIMOS Public Extragalactic Redshift Survey (VIPERS), which provides measured spectra for $\sim$ 90000 objects in two fields (W1 and W4) of the Canada-France-Hawaii Telescope
Legacy Survey Wide (CFHTLS-Wide\footnote{http://www.cfht.hawaii.edu/Science/CFHLS/}). The spectroscopic observations were carried out by the VIMOS multi-object spectrograph \citep{2003SPIE.4841.1670L} attached on the ESO Very Large Telescope (VLT). The spectra span wavelength $5500-9500 \mathrm{\AA}$ with moderate resolution ($R \simeq 220$). Combining their own WIRCam observation with the photometric data from T0007\footnote{http://www.cfht.hawaii.edu/Science/CFHLS/T0007/} release of the CFHTLS Wide photometric survey, GALEX \citep{2005ApJ...619L...1M}, and the VISTA Deep Extragalactic Observations \citep{2013MNRAS.428.1281J}, \cite{2016A&A...590A.102M} constructed a multi-band photometric catalog for the VIPERS sky footprints, called VIPERS Multi-Lambda Survey, that includes photometry at two UV bands NUV and FUV, five optical bands u, g, r, i, z, and one Near-IR bands  $K_{\mathrm{s}}$ or $K_{\mathrm{video}}$ (only for part of W1).

We need to take into account the observational effects of the VIPERS observation. Following their notions, we can decompose the observational effects of VIPERS into radial selection function and angular selection function. Firstly, the radial selection function is induced by the color target pre-selection which aims to ensure that only galaxies with redshift greater than $0.5$ are included in the parent photometric sample. This effect can be quantified by the color sampling rate ($CSR$) which was estimated by \cite{2014A&A...566A.108G} as a function of redshift. At $z<0.6$, $CSR$ can be modeled as $CSR(z) = 1/2 -1/2 \erf\left[b\left(z_{\mathrm{t}}-z\right)\right]$ where the $\erf$ is the error function, and $b=10.8$ and $z_{\mathrm{t}}=0.444$ are the best-fit parameters. At $z\geq0.6$, the survey is highly complete for $i_{\mathrm{AB}}<22.5$ (i.e. $CSR(z)=1$). As for the angular selection function, in addition to the photometric and spectroscopic masks, the target sampling rate ($TSR$) and the spectroscopic success rate ($SSR$) have been evaluated for each galaxy \citep{2018A&A...609A..84S}. The former is the probability that a galaxy in the parent sample is selected for the spectroscopic observation, and the latter accounts for the probability that the spectrum was successfully obtained. Therefore, we weight every galaxy by the inverse of $CSR$, $TSR$ and $SSR$
\begin{equation}
w^i = CSR^{-1} \times TSR^{-1} \times SSR^{-1}. \label{equ:weight}
\end{equation}

Nevertheless, as the strength of an \oii emitter depends on the color, the intrinsic relation between \oii and NUV may be biased by the  target color selection. Therefore, we restrict the galaxy sample in the range of $0.6 \leq z<1.45$ to ensure that $CSR(z)=1$. Finally, only galaxies with secure redshift determination \cite[i.e. redshift flags 2 to 9;][]{2018A&A...609A..84S} are used in our study. 

\subsection{Measurement of $L_{\oiirm}$ and $M_{\nuv}$} \label{subsec:measurement}
The flux has been fully calibrated for the spectra of VIPERS galaxies \citep{2014A&A...562A..23G}, thus we can directly use these cleaned-spectra without extra correction for slit losses. For each galaxy spectrum, we first shift it to the rest-frame and then mask the pixels in the region centered at $3727 \mathrm{\AA}$ with width $\pm 20 \mathrm{\AA}$, which accounts for the broadening of spectral line caused by the resolution $R=220$ of VIMOS spectrograph. The continuum is estimated by a sixth-order polynomial fitting in the range of $3727 \pm 200 \mathrm{\AA}$. After subtracting the continuum from the original spectra, we model the \oii doublet by a single Gaussian profile rather than two Gaussian functions because of the limited spectrum resolution. We use the Levenberg-Marquardt algorithm \citep{levenberg1944method, marquardt1963algorithm} which is an optimized least squares method to fit the subtracted spectrum and derive the best-fitting parameters as well as their covariance matrix. The flux of the \oii emission line $F_{\oiirm}$ is obtained from the best-fitting Gaussian profile, and its measurement uncertainty is estimated from the covariance matrix. In APPENDIX \ref{sec:check}, we check our \oii measurement method, by applying it to the spectra from the VIMOS VLT Deep Survey \citep[VVDS,][]{2005A&A...439..845L,2013A&A...559A..14L} and comparing our results with the \oii flux provided by \cite{2009A&A...495...53L}. There is a small systematic difference (less than $10\%$) between the two measurements, maybe due to the fact that they used a different method for the continuum fitting. Nevertheless, the two measurements agree well overall, and this slight difference does not significantly affect our subsequent conclusions. By applying our method to a total of 54166 galaxies with the redshift flag $\geq 2$ in the VIPERS sample, eventually we have detected the \oii flux for 45139 galaxies (i.e. $F_{\oiirm}>0$ with the error successfully calculated, see APPENDIX \ref{sec:check} for more details), of which 36714 have signal to noise ratios ($SNR$) higher than 5.

We correct the $F_{\oiirm}$ for the foreground dust extinction of the Milky Way by
\begin{equation}
F^{\mathrm{cor}}_{\oiirm}=F_{\oiirm}\times 10^{0.4E(B-V)k\left(\lambda^{\mathrm{obs}}_{\oiirm}\right)}, 
\end{equation}
where $E(B-V)$ is the color excess taken from the dust map provided by \cite{1998ApJ...500..525S}, and the  $k\left(\lambda\right)$ is the reddening curve  from \cite{2000ApJ...533..682C}.
Then, we calculate \oii luminosity $L_{\oiirm}$ utilizing 
\begin{equation}
L_{\oiirm} = 4\mathrm{\pi} F^{\mathrm{cor}}_{\oiirm} D^{2}_L, 
\end{equation}  
where $D_L$ is the luminosity distance defined as
\begin{equation}
D_L\left(z\right) = \left(1+z\right)\frac{c}{H_0}\int_{0}^{z}\frac{\mathrm{d} z^\prime}{\sqrt{\Omega_{m,0}\left(1+z^\prime\right)^3 + \Omega_{\Lambda,0}}}. 
\end{equation}

Simultaneously, utilizing the eight-band photometric data from the VIPERS Multi-Lambda Survey, we perform spectral energy distribution (SED) template fitting using {\tt\string LE PHARE} \citep{2002MNRAS.329..355A,2006A&A...457..841I} to all galaxies and derive their near-ultraviolet band absolute magnitudes $M_{\nuv}$ as well as other physical parameters such as stellar mass, age and SFR. In our SED fitting, the stellar population synthesis models are taken from the library provided by \cite{2003MNRAS.344.1000B} and the \cite{2003PASP..115..763C} initial stellar mass function is adopted. We set three metallicities $0.4Z_{\sun}$, $1Z_{\sun}$ and $2.5Z_{\sun}$ and consider a delayed star formation
history (SFH) $\sim t \exp\left(-t/ \tau \right)$ where the timescale $\tau$ uniformly spans in the logarithm space from $10^7\mathrm{yr}$ to $1.258 \times 10^{10}\mathrm{yr}$. As for the dust extinction, the color excess $E(B-V)$ is taken from 0 to 0.5 and the starburst reddening curve \citep{2000ApJ...533..682C} is applied to calculate the attenuation factor.

Figure \ref{fig:L_M_obs_2d} shows the two-dimensional joint probability density distribution (PDF) of $M_{\nuv}$ and $\log L_{\oiirm}$ for galaxies with $SNR> 5$ in different redshift bins. The color bars display the value of this joint PDF that is calculated based on the kernel density
estimate (KDE) with a Gaussian kernel. Clearly, the intensity of \oii is tightly correlated with $M_{\nuv}$. This phenomenon is expected, as galaxies with stronger UV radiation are more likely to photo-ionize neutral oxygen and produce stronger \oii emission line.

\begin{figure*}
	\centering
	\includegraphics[scale=0.7]{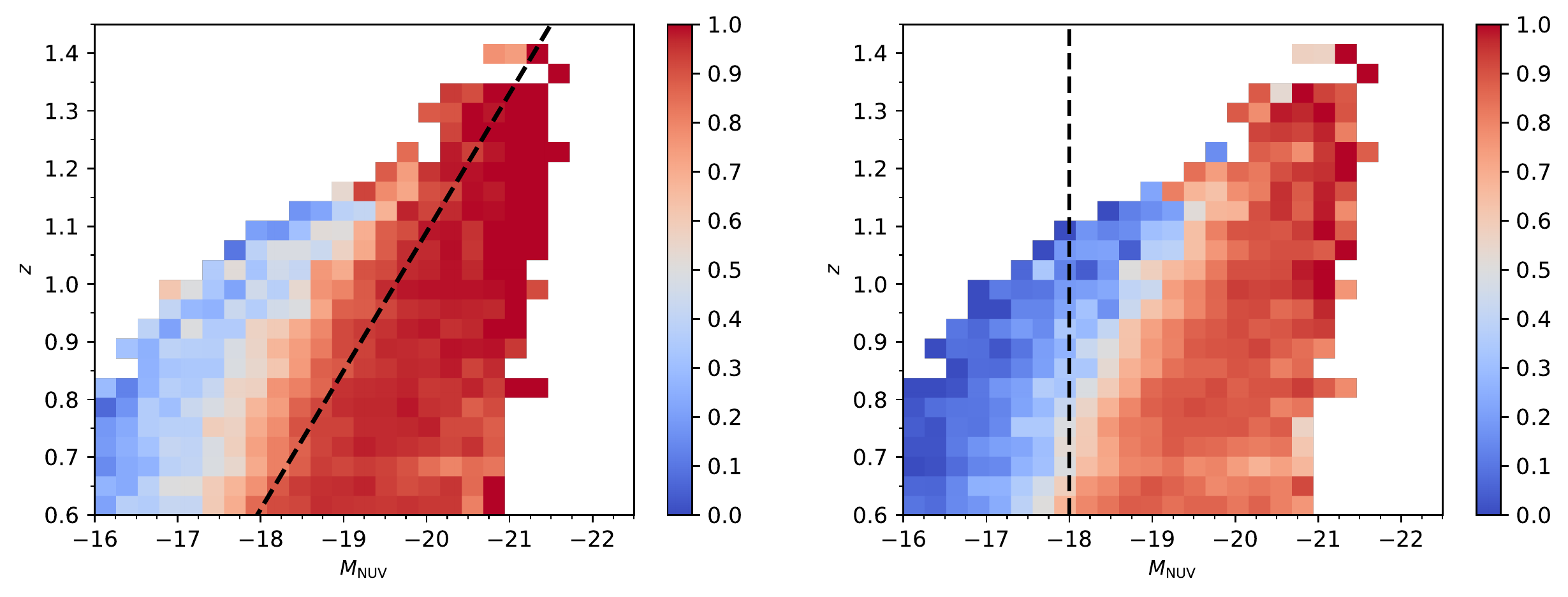}
	\caption{The measurement success rate ($MSR$) for \oii emission lines. The left panel (for method 1) shows $MSR^{SNR>1} = N^w(SNR>1)/N^w$, which is the ratio of the weighted number of galaxies with $L_{\oiirm}$ $SNR>1$ to all galaxies in each $M_{\nuv}$ and redshift bin. Similar to the left panel, the right panel (for method 2) displays $MSR^{SNR>5} = N^w(SNR>5)/N^w$. The color bar represents the value of the $MSR$ at each bin. Both panels only show the bins each containing more than 20 galaxies. The two black dashed lines denote the $M_{\nuv}$ cut for the two methods. For method 1, we assume the galaxies falling into the region right of the black line have complete $L_{\oiirm}$ measurement, and we use them for our modeling. Whereas, for method 2, we use all the galaxies in the region right of the black line ($M_{\nuv}<-18$), regardless of their $SNR$, but we mask out those $L_{\oiirm}$ with $SNR<5$ in our fitting approach.  \label{fig:MeasureSuccessRate2d}}
\end{figure*}

\begin{deluxetable*}{ccccccccc}
	\tablenum{1}
	\tablecaption{The best-fitting model parameters with their $1\sigma$ uncertainties, and the number of galaxies used in the two methods. \label{tab:result}}
	\tablehead{
		\colhead{Method} & \colhead{Redshift Range} & \colhead{$N_{\mathrm{galaxy}}$} & \colhead{$N^{SNR>1}_{\mathrm{galaxy}}$} & \colhead{$N^{SNR>5}_{\mathrm{galaxy}}$} &
		\colhead{$\alpha$} & \colhead{$\omega$} & \colhead{$k$} & \colhead{$b$}
	}
	\startdata
	{        }&$0.6\leq z<0.7$& \nodata& 10298&\nodata& $-2.4719^{+0.0453}_{-0.0460}$&$0.3772^{+0.0026}_{-0.0026}$&$-0.4$&$34.0436^{+0.0024}_{-0.0024}$\\
	{        }&$0.7\leq z<0.8$& \nodata& 8537&\nodata& $-2.4270^{+0.0475}_{-0.0498}$&$0.3704^{+0.0028}_{-0.0028}$&$-0.4$&$34.0672^{+0.0027}_{-0.0027}$\\
	{Method 1}&$0.8\leq z<0.9$& \nodata&5923&\nodata&$-2.5072^{+0.0587}_{-0.0599}$&$0.3665^{+0.0032}_{-0.0032}$&$-0.4$&$34.0733^{+0.0030}_{-0.0030}$\\
	{        }&$0.9\leq z<1.0$&\nodata&4011&\nodata&$-2.7083^{+0.0793}_{-0.0788}$&$0.3514^{+0.0036}_{-0.0036}$&$-0.4$&$34.0713^{+0.0033}_{-0.0034}$\\
	{        }&$1.0\leq z<1.1$&\nodata&1757&\nodata&$-2.0849^{+0.0883}_{-0.0903}$&$0.3358^{+0.0052}_{-0.0054}$&$-0.4$&$34.1133^{+0.0054}_{-0.0056}$\\
	{        }&$0.6\leq z<1.45$&\nodata&31709&\nodata&$-2.4357^{+0.0250}_{-0.0253}$&$0.3678^{+0.0014}_{-0.0014}$&$-0.4$&$34.0695^{+0.0013}_{-0.0013}$\\
	\hline
	{        }&$0.6\leq z<0.7$& 11799& \nodata&9596& $-2.6805^{+0.0454}_{-0.0465}$&$0.4341^{+0.0031}_{-0.0030}$&$-0.4$&$34.0518^{+0.0025}_{-0.0025}$\\
	{        }&$0.7\leq z<0.8$& 11062& \nodata&9045& $-2.5538^{+0.0474}_{-0.0473}$&$0.4226^{+0.0031}_{-0.0032}$&$-0.4$&$34.0800^{+0.0026}_{-0.0027}$\\
	{Method 2}&$0.8\leq z<0.9$& 8679&\nodata&6873&$-2.0802^{+0.0485}_{-0.0482}$&$0.4205^{+0.0040}_{-0.0040}$&$-0.4$&$34.0741^{+0.0037}_{-0.0037}$\\
	{        }&$0.9\leq z<1.0$&6113&\nodata&4785&$-2.0293^{+0.0586}_{-0.0587}$&$0.3989^{+0.0046}_{-0.0046}$&$-0.4$&$34.0629^{+0.0042}_{-0.0044}$\\
	{        }&$1.0\leq z<1.1$&2791&\nodata&2128&$-1.2168^{+0.0790}_{-0.0785}$&$0.3572^{+0.0078}_{-0.0078}$&$-0.4$&$34.0660^{+0.0098}_{-0.0104}$\\
	{        }&$0.6\leq z<1.45$&42392&\nodata&34013&$-2.1387^{+0.0216}_{-0.0218}$&$0.4168^{+0.0017}_{-0.0017}$&$-0.4$&$34.0732^{+0.0015}_{-0.0015}$
	\enddata
\end{deluxetable*}

\begin{figure*}
	\centering
	\includegraphics[scale=0.7]{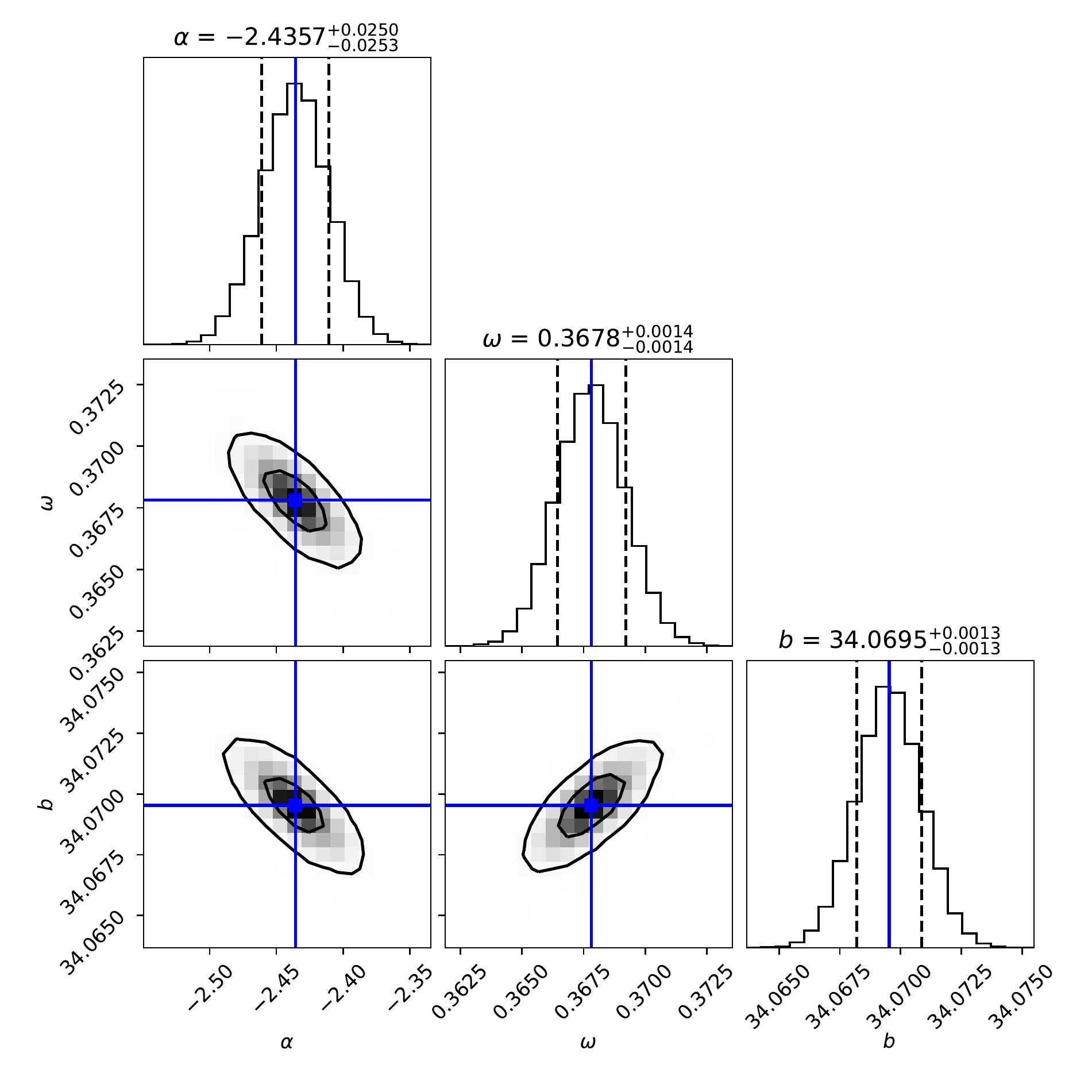}
	\caption{The joint posterior probability distribution for model parameters as well as their marginal probability distribution obtained by method 1 in the entire redshift range $0.6\leq z<1.45$. The best-fit values of $\alpha$, $\omega$ and $b$ are labeled as blue solid lines. The black dashed lines denote the $16\%$ and $84\%$ percentile of the marginal distribution. The confidence intervals of $68\%$  $(1\sigma)$ and $95\%$ $(2\sigma)$ are represented by the internal and external contours, respectively. \label{fig:mcmc}}
\end{figure*}

\begin{figure*}
	\centering
	\includegraphics[scale=0.6]{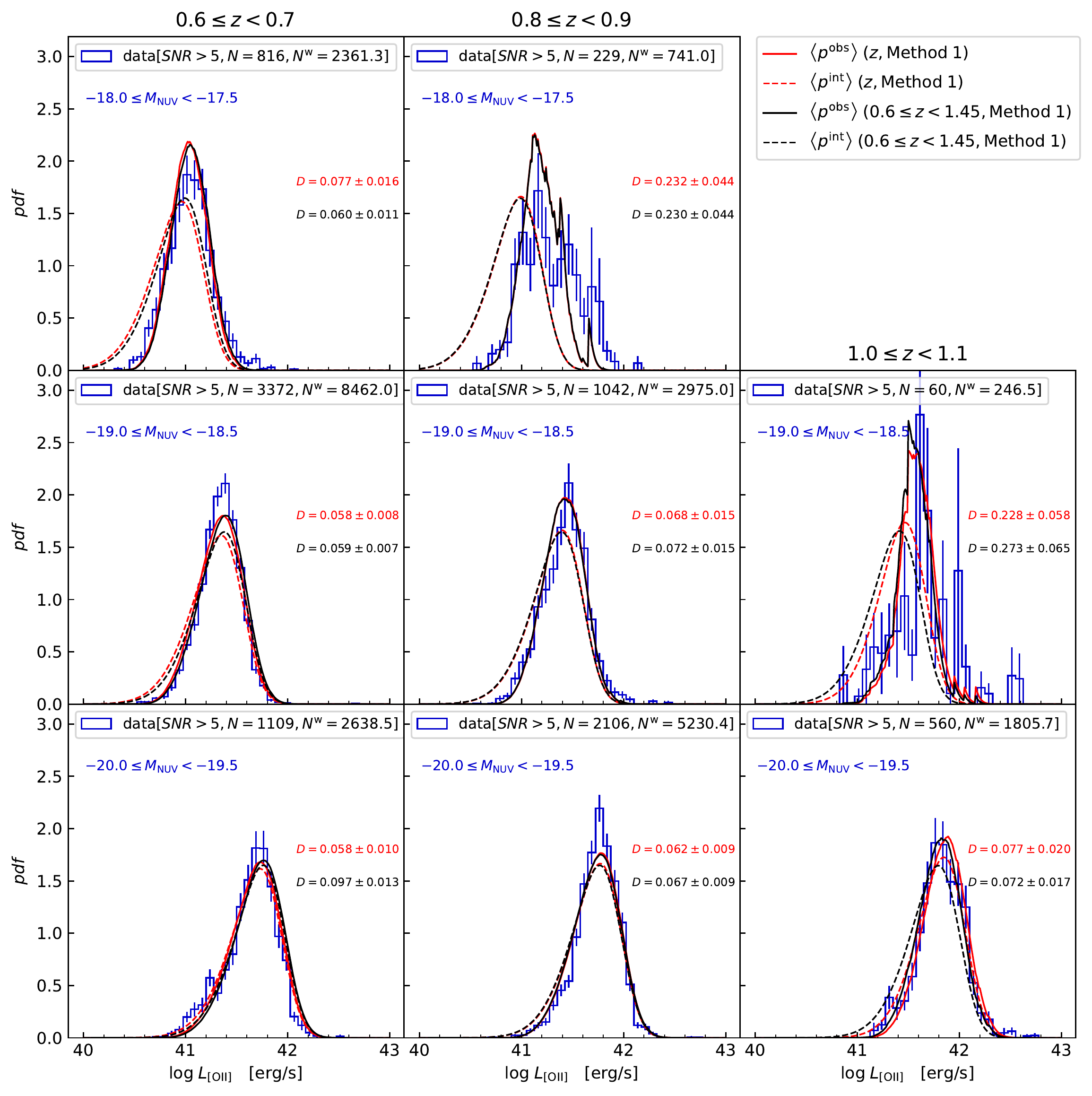}
	\caption{The observed $\log L_{\oiirm}$ distribution of galaxies with $SNR>5$ versus the model predictions with best-fit parameters derived by the method 1. The blue histograms with Poisson errors in each panel show the observed $\log L_{\oiirm}$ distribution. Different panels in the same column show the distributions for different $M_{\nuv}$ at the same redshift. The solid lines represent our model $\left \langle p^{\mathrm{obs}}\right\rangle$ calculated by Equation \ref{equ:ave_obs} for  $SNR>5$. For comparison, the average intrinsic distribution $\left \langle p^{\mathrm{int}}\right\rangle$ defined by Equation \ref{equ:ave_int} are plotted as dashed lines. We plot the model predictions with the best-fit parameters not only from the individual $z$ bin itself but also from the entire redshift range $0.6\leq z < 1.45$. Here the $D$, whose color corresponds to the color of the solid lines, is Kolmogorov-Smirnov (K-S) statistic that is used to quantify the maximum (supremum) distance between the CDF of the predicted distribution and the empirical cumulative distribution function (ECDF) of the ordered observational data. \label{fig:histogram_method1}}
\end{figure*}

\begin{figure*}
	\centering
	\includegraphics[scale=0.6]{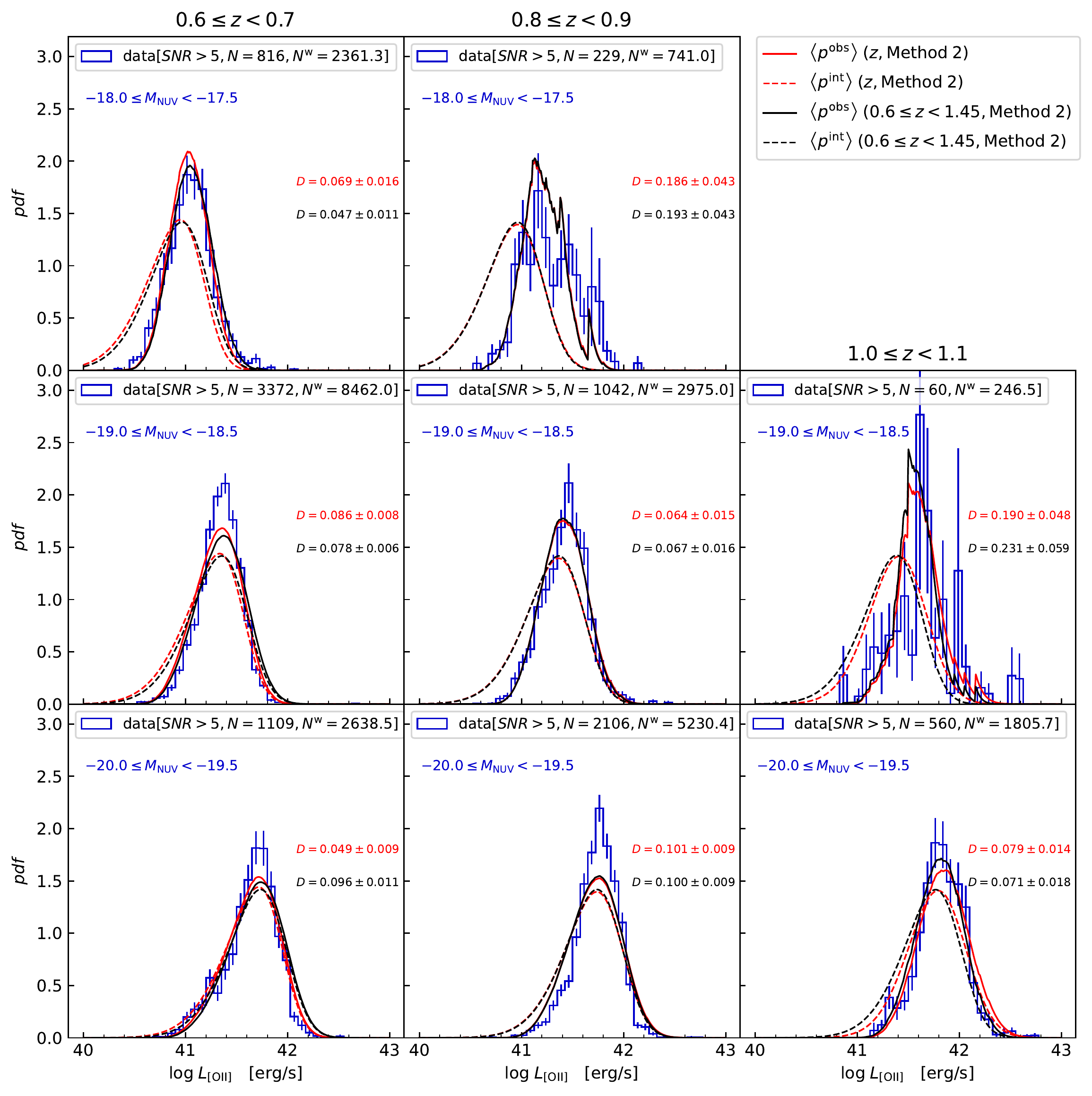}
	\caption{Same as Figure \ref{fig:histogram_method1}, but the $\left \langle p^{\mathrm{obs}}\right\rangle$ and $\left \langle p^{\mathrm{int}}\right\rangle$ are predicted with best-fit parameters derived by the method 2. \label{fig:histogram_method2}}
\end{figure*}

\begin{figure*}
	\centering
	\includegraphics[scale=0.8]{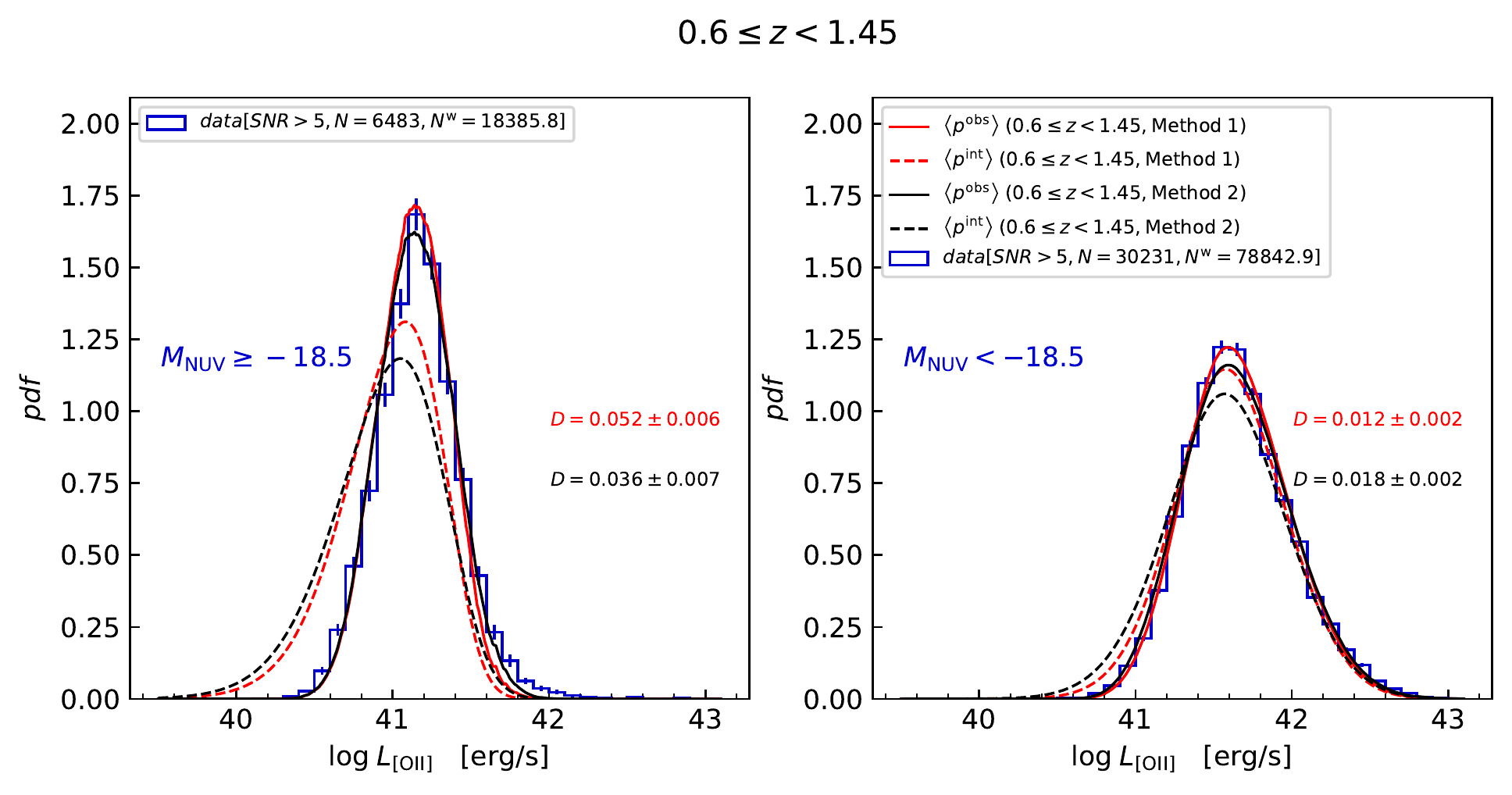}
	\caption{Same as Figure \ref{fig:histogram_method1} and Figure \ref{fig:histogram_method2}, but we display the observed $L_{\oiirm}$ distribution in the entire redshift range $0.6\leq z < 1.45$. The left panel shows the distribution for the faint subsample ($M_{\nuv}\geq-18.5$) and the right panel displays the one for the luminous subsample ($M_{\nuv}<-18.5$). The $\left \langle p^{\mathrm{obs}}\right\rangle$ and $\left \langle p^{\mathrm{int}}\right\rangle$ are predicted with best-fit parameters derived by the both methods. \label{fig:histogram_compare12}}
\end{figure*}

\section{Intrinsic Conditional Distribution Function} \label{sec:methodology}
In this section, we will describe the methods we use to model the intrinsic conditional probability density distribution $p^{\mathrm{int}}\left(L_{\oiirm}|M_{\nuv}\right)$ from the observational data. We note that because of the given sensitivity, the observation may fail in yielding a \oii line detection for a relatively weaker emitter. This incompleteness depends on the strength and on the redshift of the emitters.  We will pay particular attention to the incompleteness effect of the \oii emission line measurement. Two methods are adopted to overcome the observational effects, and, as will be shown, produce similar results.

\subsection{Intrinsic model} \label{subsec:Intrinsic}
When predicting \oii LF from the NUV LF, the most critical step is to derive the intrinsic relation between $L_{\oiirm}$ and $M_{\nuv}$. 

As shown in Figure \ref{fig:L_M_obs_2d}, the mean relationship of $M_{\nuv}$ and $\log L_{\oiirm}$ can be reasonably described via a linear model. Additionally, at a given $M_{\nuv}$, the dispersion of the $\log L_{\oiirm}$ distribution is less than 1 dex. Considering the above two factors, we attempt to construct a linear model plus a skew-normal distribution \citep{o1976bayes} for $p^{\mathrm{int}}\left(L_{\oiirm}|M_{\nuv}\right)$ that can be expressed as

\begin{eqnarray}
p^{\mathrm{int}}\left(L_{\oiirm}|M_{\nuv}\right) =&\frac{2}{\omega}&\phi\left(\frac{\log L_{\oiirm-\xi}}{\omega}\right) \nonumber \\ &\times&  \Phi\left(\alpha\left(\frac{\log L_{\oiirm-\xi}}{\omega}\right)\right), \label{equ:model}
\end{eqnarray}
where $\phi(x)=\frac{1}{\sqrt{2\pi}}\mathrm{e}^{-\frac{x^2}{2}}$ denote the standard normal PDF and $\Phi(x)=\int_{-\infty}^{x}\phi(t)\,\mathrm{d}t$ is defined as the cumulative distribution function (CDF). The parameter $\alpha$ ($\alpha = 0$ for a normal distribution) and $\omega$ describe the skewness and scale of the skew-normal PDF respectively, the location parameter $\xi = kM_{\nuv} + b$ indicates the linear relation between $M_{\nuv}$ and $\log L_{\oiirm}$. It should be noticed that the location parameter $\xi$ is different from the location $\xi_{\mathrm{max}}$ where the PDF reaches the maximum value. The latter is defined as the mode of this distribution and can also be regarded as the expectation value of $\log L_{\oiirm}$ at given $M_{\nuv}$. It can be numerically calculated through 
\begin{equation}
\log L^{\mathrm{exp}}_{\oiirm}=\xi_{\mathrm{max}} = kM_{\nuv} + b + \omega m_0(\alpha), \label{equ:xi}
\end{equation}
where $m_0({\alpha})$ is accurately approximated \citep{azzalini2013skew} as
\begin{eqnarray}
m_0(\alpha) = \mu_z - \frac{\gamma_1 \sigma_z}{2} - \frac{\sgn (\alpha) }{2}\mathrm{e}^{-\frac{2\pi}{\left\vert \alpha \right\vert}} \label{equ:m0}
\end{eqnarray}
with
\begin{eqnarray}
\delta &=& \alpha/\sqrt{1+\alpha^2}.\nonumber\\
\mu_z &=& \sqrt{2/\pi}\delta, \sigma_z = \sqrt{1-\mu_z^2} \nonumber \\
\gamma_1 &=& \frac{4-\pi}{2}\frac{\left(\delta \sqrt{2/\pi}\right)^3}{\left(1-2\delta^2/\pi\right)^{3/2}} \nonumber
\end{eqnarray}
In principle, the intrinsic distribution $p^{\mathrm{int}}\left(L_{\oiirm}|M_{\nuv}\right)$ may be described by other suitable statistical models. The skew-normal distribution is just one concise model that is simple and can describe the intrinsic distribution well.

\subsection{Method 1} \label{subsec:method1}
In order to derive the intrinsic model for $p^{\mathrm{int}}\left(L_{\oiirm}|M_{\nuv}\right)$, we must properly take into account the incompleteness introduced by the $L_{\oiirm}$ measurement. Namely, we should ensure that the $L_{\oiirm}$ data is complete for a certain $M_{\nuv}$. Nevertheless, for a part of \oii emitters with weak emission lines or noisy spectra, it is difficult to derive their emission line profiles successfully, and thus these galaxies have low $SNR$s of $L_{\oiirm}$ measurement. Therefore, we regard a measurement of $SNR>1$ as a meaningful one only.

For the purpose of overcoming this selection effect, in the first method (method 1) we attempt to construct a complete sample for $L_{\oiirm}$ emitters. In analog to defining $TSR$ and $SSR$, we define the measurement success rate ($MSR$) as the probability that the line flux is successfully measured for a galaxy of given $M_{\nuv}$ at redshift $z$.  We divide the galaxies into two-dimensional bins according to their redshift and $M_{\nuv}$, and then compute the $MSR^{SNR>1} = N^{\mathrm{w}}(SNR>1)/N^{\mathrm{w}}$ as the ratio of the weighted number of galaxies with $SNR>1$ to all galaxies in that bin. We display the $MSR^{SNR>1}$ as a function of $M_{\nuv}$ and redshift $z$ in the left panel of Figure \ref{fig:MeasureSuccessRate2d}, where only the bins containing more than 20 galaxies are plotted. Apparently, at a given redshift, the $MSR$ increases with the NUV brightness to a critical value $M^{\mathrm{c}}_{\nuv}$, and then stabilizes between 0.9 and 1 at $<M^{\mathrm{c}}_{\nuv}$. Furthermore, the $MSR$ also decreases as the redshift increases at a given $M_{\nuv}$.  The black dashed line approximately represents the critical value $M^{\mathrm{c}}_{\nuv}$ dividing the plot into two regions: in the right region (right of the dividing line) the  $F_{\oiirm}$ measurement is highly complete, with $MSR$ being greater than 0.9 at almost all grid points and reaching 0.95 overall in the region. In the left region the measurement could be very incomplete, the effect of which must be taken into account in the analysis. It should be noted that the measured $F_{\oiirm}$ with low $SNR$ may be fairly inaccurate even in the complete region. In other words, some galaxies with weak enough intrinsic \oii luminosities may enter into the sample of $SNR > 1$ due to the photon noise in the measurements, and vice versa. To overcome the impurity and incompleteness effects, we must take the measurement uncertainties into consideration in our fitting process. As a result, for the {\it i}-th galaxy in our sample, we assume the observed $L^i_{\oiirm}$ follow a Gaussian distribution for the given uncertainty $\sigma^i _{L_{\oiirm}}$, and convolve this error distribution with the intrinsic PDF $p^{\mathrm{int}}\left(L^i_{\oiirm}|M^i_{\nuv}\right)$ to account for the observed PDF  
\begin{eqnarray}
p^{\mathrm{obs}}&\left(L^i_{\oiirm}|\theta,M^i_{\nuv},\sigma^i _{L_{\oiirm}}\right) =\int p^{\mathrm{int}}\left(L^{\prime}_{\oiirm}|\theta,M^i_{\nuv}\right)\nonumber \\ 
&\times N\left(L^{\prime}_{\oiirm}-L^{i}_{\oiirm}|\sigma^i _{L_{\oiirm}}\right) \,\mathrm{d}L^{\prime}_{\oiirm},
\end{eqnarray}
where the $\theta$ are the parameters of the intrinsic distribution (Equation \ref{equ:model}).
Consequently, the logarithmic likelihood function $\ln \mathcal{L}^{\mathrm{M1}}$ for the first method M1 can be written as
\begin{eqnarray}
\ln \mathcal{L}^{\mathrm{M1}} = \sum_{i} w^i \ln p^{\mathrm{obs}}&\left(L^i_{\oiirm}|\theta,M^i_{\nuv},\sigma^i _{L_{\oiirm}}\right) \label{equ:ln1}
\end{eqnarray}
where the $w^i$ is the weight of the {\it i}-th galaxy (Equation \ref{equ:weight}).

According to the Bayesian statistical theory, the posterior probability of the model parameters is proportional to the product of the likelihood function and the prior probability
\begin{eqnarray}
p\left(\theta|L_{\oiirm}, M_{\nuv}, \sigma _{L_{\oiirm}}\right) \propto&
\mathcal{L}&\left(L_{\oiirm}|\theta,M_{\nuv},\sigma _{L_{\oiirm}}\right) \nonumber \\ &\times& p\left(\theta|M_{\nuv},\sigma _{L_{\oiirm}}\right).
\end{eqnarray}
Moreover, in order to investigate the redshift evolution of the intrinsic distribution, we divide the data into five redshift bins: $0.6\leq z<0.7$, $0.7\leq z<0.8$, $0.8\leq z<0.9$, $0.9\leq z<1.0$, $1.0\leq z<1.1$. The number of galaxies we used to fit in each redshift bin are shown in Table \ref{tab:result}. Considering the proportional relationship between the luminosity of the NUV radiation and the luminosity of the \oii emission line, we fix the slope $k$ as $-0.4$. In fact, we have tried to set $k$ as a free parameter in the fitting, but we find that $k$ is indeed close to $-0.4$ in all redshift bins. We choose the flat prior distributions for the other three parameters: $-5<\alpha<5$, $0.1<\omega<1$, $0<b<50$.

We utilize the Markov Chain Monte Carlo (MCMC) approach to explore the space of the three parameters to obtain their posterior probability. A {\tt\string python} package called as {\tt\string emcee} \citep{2013PASP..125..306F} is used to perform the MCMC sampling in the parameter space. We randomly assign initial positions for 30 chains and run each chain by 5000 steps. The first 300 steps (about 10 times the integrated auto-correlation time) of each chain are discarded to ensure the convergence of the MCMC samples. In Table \ref{tab:result}, we show the fitting results of these parameters as well as their $1\sigma$ error. We find that the parameters do not change significantly with the redshift in the range of $0.6<z<1.1$. This is very encouraging, as it implies that the parameters do not change either with the NUV luminosity because the sample contains more NUV luminous galaxies at a higher redshift.  Therefore, we have also analyzed for the entire sample of galaxies within the redshift $0.6\leq z<1.45$, and list their model parameters in Table \ref{tab:result}.  We display the joint posterior probability distribution for any two of the parameters $\alpha$, $\omega$ and $b$ as well as the marginalized probability distribution for a single parameter in Figure \ref{fig:mcmc} obtained by method 1 for the entire sample.

Let us first check the mean relationship of $M_{\nuv}$ and $L_{\oiirm}$, which can be constructed with parameters $k$ and $b$. For the purpose of exploring the redshift evolution of this relationship, using the Equation \ref{equ:xi} and Equation \ref{equ:m0}, we numerically compute the $\log L^{\mathrm{exp}}_{\oiirm}$ of the $L_{\oiirm}$ PDF at given $M_{\nuv}$ with the best-fit parameters in each redshift bin and show it in Figure \ref{fig:L_M_obs_2d} as a solid line. Meanwhile, the $\log L^{\mathrm{exp}}_{\oiirm}$ for the entire redshift range $0.6\leq z <1.45$ is also shown in the other panels as a dotted line. Although the best-fit $b$ is slightly different at various redshifts, the $\log L^{\mathrm{exp}}_{\oiirm}$ does not indicate a significant redshift evolution trend. This fact suggests that the mean relationship of $M_{\nuv}$ and $L_{\oiirm}$ is nearly redshift-independent and the assumption that $k=-0.4$ is reasonable. 

Now let us further discuss about the distribution functions. In Figure \ref{fig:histogram_method1}, we display the observed PDF of $L_{\oiirm}$ for galaxies with the $SNR>5$ according to their $M_{\nuv}$ and $z$ (histograms). We have used the weight $w^i$ (Equation \ref{equ:weight}) to calculate the PDF and the Poisson error bars. In order to compare our best-fit intrinsic distribution function $p^{\mathrm{int}}$ with the observational histogram, we calculate the weighted average observed PDF $\left \langle  p^{\mathrm{obs}}\right\rangle$ by
\begin{eqnarray}
\left \langle p^{\mathrm{obs}}\right\rangle = \frac{\sum_{i}w^i p^{\mathrm{obs,cod}}\left(L_{\oiirm}|M^i_{\nuv},\sigma^i _{L_{\oiirm}}\right)}{\sum_{i}w^i} \label{equ:ave_obs},
\end{eqnarray}
where the $ p^{\mathrm{obs,cod}}$ expressed as 
\begin{eqnarray}
&p^{\mathrm{obs,cod}}&\left(L_{\oiirm}|M^i_{\nuv},\sigma^i _{L_{\oiirm}}\right) \nonumber \\
&=&  \frac{p^{\mathrm{obs}}\left(L_{\oiirm}|M^i_{\nuv},\sigma^i _{L_{\oiirm}}\right)}{1-C^{\mathrm{obs}}\left(L^i_{\oiirm,\mathrm{thr}}|M^i_{\nuv},\sigma^i _{L_{\oiirm}}\right)}
\end{eqnarray}
is the conditional PDF for the cut of $SNR>5$ and $C^{\mathrm{obs}}$ expressed as 
\begin{eqnarray}
&C^{\mathrm{obs}}&\left(L^{i}_{\oiirm,\mathrm{thr}}|\theta,M^i_{\nuv},\sigma^i _{L_{\oiirm}}\right) \nonumber \\
&=& \int^{L^{i}_{\oiirm},\mathrm{thr}} p^{\mathrm{obs}}\left(L^{\prime}_{\oiirm}|\theta,M^i_{\nuv},\sigma^i _{L_{\oiirm}}\right)\,\mathrm{d} L^{\prime}_{\oiirm} \nonumber \\ \label{equ:Cobs}
\end{eqnarray}
represents the cumulative probability function (CDF) of the observational distribution $p^{\mathrm{obs}}$ below the \oii $SNR$ threshold ($L^{i}_{\oiirm,\mathrm{thr}}=5\sigma^i _{L_{\oiirm}}$ for the {\it i}-th galaxy).

 Obviously, with $\left \langle  p^{\mathrm{obs}}\right\rangle$ we have properly accounted for the incompleteness of measuring the \oii line flux for given redshift and $M_{\nuv}$. The fitted curves $\left \langle  p^{\mathrm{obs}}\right\rangle$ based on the method 1 overall match well with the observed ones. The parameters we obtained for the entire sample of $0.6 \leq z <1.45$ also fit with the subsample at different redshift. To make a statistical comparison, we tried to assess this method and (or) possible redshift dependence of the parameters by taking  the non-parametric  Kolmogorov-Smirnov (K-S) test \citep{kolmogorov1933sulla, smirnov1948table}. We calculate the empirical cumulative distribution function (ECDF) of the ordered observational data, and compare it to the CDF of the distribution $\left \langle  p^{\mathrm{obs}}\right\rangle$. The maximum (supremum) distance between the ECDF and CDF is defined as the K-S statistic $D$. To evaluate the impact of sample size, we estimate the $1\sigma$ uncertainty of $D$ by the bootstrap re-sampling method and annotate their values in each panel of Figure \ref{fig:histogram_method1}. For one observed distribution (one panel of Figure \ref{fig:histogram_method1}), the statistic $D$ for a single redshift $\left \langle  p^{\mathrm{obs}}\right\rangle (z)$ is close to that for the entire redshift range $\left \langle  p^{\mathrm{obs}}\right\rangle (0.6 \leq z <1.45)$, thus we do not find any significant redshift dependence of the parameters.  

It would be interesting to compare the intrinsic distribution  $p^{\mathrm{int}}$ with the $\left \langle  p^{\mathrm{obs}}\right\rangle$. Considering the range of $M_{\nuv}$ of the galaxies at each redshift bin, we calculate the average intrinsic PDF $\left \langle p^{\mathrm{int}}\right\rangle$ by
\begin{eqnarray}
\left \langle p^{\mathrm{int}}\right\rangle = \frac{\sum_{i} w^i p^{\mathrm{int}}\left(L_{\oiirm}|M^i_{\nuv}\right)}{\sum_{i}w^i}  \label{equ:ave_int},
\end{eqnarray}
and we have plotted them as the dashed lines in Figure \ref{fig:histogram_method1}. The figure shows that the faint subsample ($M_{\nuv}\geq-18.5$) are very incomplete at all redshift, and the brighter ones become increasingly incomplete with redshift. This indicates how important it is to correct for the incompleteness in deriving the intrinsic distributions. Clearly, for both the $\left \langle p^{\mathrm{int}}\right\rangle$ and $\left \langle p^{\mathrm{obs}}\right\rangle$, the PDF for the entire redshift range $0.6\leq z< 1.45$ is close to that for each single redshift bin. It means that not only the mean relationship of $M_{\nuv}$ and $L_{\oiirm}$ but also the intrinsic scatter distribution of $L_{\oiirm}$ at given $M_{\nuv}$ is nearly redshift independent.

\subsection{Method 2} \label{subsec:method2}
In our method 1, although we have ensured  the completeness of \oii flux measurement at $\sim$ 95 \% level and have also corrected for the influence of measurement errors, we have had to adopt the critical value $M^{\mathrm{c}}_{\nuv}$ that limits the number of galaxies we can use. In order to check whether our results are robust to the selection of the complete sample, we develop the second method (method 2) to account for the incompleteness in a different way. We also use the $MSR$, but here we calculate $MSR^{SNR>5} = N^{\mathrm{w}}(SNR>5)/N^{\mathrm{w}}$ that is the probability of measuring $L_{\oiirm}$ with high $SNR >5$, and display it in the right panel of Figure \ref{fig:MeasureSuccessRate2d}. We assume that those \oii lines with $SNR>5$ can be 100\% successfully measured. In the opposite, for the remaining galaxies with $SNR<5$, we regard their $L_{\oiirm}$ as unsuccessfully measured. However, we notice that $MSR^{SNR>5}$ becomes very low for those faint galaxies ($M_{\nuv}>-18$), which may make our fitting results unstable. For instance, the $MSR^{SNR>5}$ at $M_{\nuv}=-16$ is less than 0.3 at any redshift, namely, at least 70 \% galaxies have not been measured in their $L_{\oiirm}$ fluxes. For this reason, in our fitting process, we make a loose $M_{\nuv}$ cut criteria ($M_{\nuv}<-18$) and plotted it as the black dashed line in the right panel of Figure \ref{fig:MeasureSuccessRate2d}, which enables us to avoid censoring too much data. It should be emphasized that we use {\it all} the galaxies in the region right to this dividing line, i.e. we have not only included the galaxies whose \oii fluxes have been measured successfully ($SNR>5$), but also the ones with $SNR$ below the threshold $SNR =5$ or even with negative $SNR < 0$. In statistics, the concept of this kind of model regression problem in which the data is censored (true value is unknown but the dividing threshold is known) is firstly proposed by \cite{tobin1958estimation}. \cite{hartley1985maximum} summarize this regression problem and show its application in the maximum likelihood estimation (MLE). Referring to this MLE method, we construct the logarithmic likelihood function $\ln \mathcal{L}^{\mathrm{M2}}$ for the second method 
\begin{eqnarray}
&\ln& \mathcal{L}^{\mathrm{M2}} \nonumber \\
 &=& \sum_{i} \left[ \left(1-f^i\right)w^i\ln C^{\mathrm{obs}}\left(L^{i}_{\oiirm,\mathrm{thr}}|\theta,M^i_{\nuv},\sigma^i _{L_{\oiirm}}\right) \right. \nonumber \\
 & &+ \left. f^iw^i\ln p^{\mathrm{obs}}\left(L^i_{\oiirm}|\theta,M^i_{\nuv},\sigma^i _{L_{\oiirm}}\right) \right], \label{equ:ln2}
\end{eqnarray}
where $f^i$ is a step function indicating whether its $L^i_{\oiirm}$ measurement is successful,
\begin{eqnarray}
f^i(SNR) = \left\{ \begin{array}{rcl} 0, & \mbox{if}& SNR<5 \\ 1,  & \mbox{if} & SNR\geq 5 \end{array}\right.,
\end{eqnarray}
$w^i$ is the weighting factor and $C^{\mathrm{obs}}$ defined by Equation \ref{equ:Cobs} represents the CDF at $5\sigma$ threshold ($L^{i}_{\oiirm,\mathrm{thr}}=5\sigma^i _{L_{\oiirm}}$).

We use the same MCMC fitting scheme as described in Section \ref{subsec:method1} for method 2 and show the best-fit parameters in Table \ref{tab:result}. Analogous to method 1, the $\log L^{\mathrm{exp}}_{\oiirm}$ derived from method 2 is plotted as the black line in Figure \ref{fig:L_M_obs_2d}.  The comparisons between the observational distributions and the model predictions $\left \langle p\right\rangle$ for method 2 are displayed in Figure \ref{fig:histogram_method2}.

From Figure \ref{fig:L_M_obs_2d}, comparing the results from method 1 and method 2, we find the corresponding $\log L^{\mathrm{exp}}_{\oiirm}$ are very close to each other in the low redshift range $0.6\leq z <0.8$ where the measurement of $L_{\oiirm}$ is relatively complete. For the higher redshift range $0.8\leq z < 1.1$, $\log L^{\mathrm{exp}}_{\oiirm}$ of method 2 shows a lower intercept, maybe due to the fact that method 2 uses more galaxies that have $SNR < 1$ which were otherwise excluded in method 1 and makes the $\log L^{\mathrm{exp}}_{\oiirm}$ shift to a lower luminosity. Additionally, the $L^{\mathrm{exp}}_{\oiirm}$ of method 2 appears to be more stable at different $z$, even for the highest redshift range $1.0\leq z <1.1$ where the $\log L^{\mathrm{exp}}_{\oiirm}$ still agrees very well with the one derived from the entire
sample ($0.6\leq z <1.45$). It may indicate that the method 2 can better correct for the bias caused by the incomplete measurements and recover the intrinsic $\log L^{\mathrm{exp}}_{\oiirm}$. But in any case, the difference between the intercepts obtained from the two methods and (or) for the different redshift bins is always less than 0.1 dex (see also Table \ref{tab:result}), indicating that the mean relation between the luminosity of $L_{\oiirm}$ and NUV is both nearly redshift independent.

We further compare the distribution functions obtained by the two methods in Figure \ref{fig:histogram_compare12}, where we show the observed $L_{\oiirm}$ distribution and predicted $\left \langle p\right\rangle$ in the entire redshift range $0.6 \leq z <1.45$. On one hand, compared to the method 1, the intrinsic distribution $\left \langle p^{\mathrm{int}}\right\rangle$ for method 2 tends to be more extended towards the low $L_{\oiirm}$ end. This is exactly what we expected, since we use more galaxies with weak or undetected \oii emission line in method 2. Besides, although the best-fit parameters $\alpha$ and $\omega$ for the two methods have obvious divergence, their observed distributions $\left \langle p^{\mathrm{obs}}\right\rangle$ are relatively consistent in general. This characteristic can be interpreted as the degeneracy of $\alpha$ and $\omega$, the absolute value of $\alpha$ increases with $\omega$ but the shape of the skew-normal distribution does not change much. On the other hand, the $\left \langle p^{\mathrm{obs}}\right\rangle$ from the method 1 seems to be closer to the observation than that from method 2 in the luminous bin ($M_{\nuv}<-18.5$), but in the faint bin ($M_{\nuv}\geq-18.5$) where the $L_{\oiirm}$ measurement is more incomplete, the method 2 shows better results because it utilizes extra information from the $L_{\oiirm}$ measurements with $SNR < 1$ compared to method 1. In general, both methods can yield a universal distribution function that can reproduce the observed $L_{\oiirm}$ distributions regardless of their redshift. Therefore, combining the NUV LF at a certain redshift, we can use the intrinsic conditional distribution functions $p^{\mathrm{int}}\left(L^i_{\oiirm}|M^i_{\nuv}\right)$ to predict the \oii LF at that redshift.

Furthermore, recently \cite{2020MNRAS.497.5432F} adopts three different semi-analytical models (SAMs) running on the MultiDark2 simulation \citep{10.1093/mnras/stw248} to study the properties of mocked \oii emitters. They have established the linear scaling relations between $L_{\oiirm}$ and the observed-frame absolute magnitude $M_{\mathrm{u}}$ and $M_{\mathrm{g}}$ at $z=0.94$, where the observed u and g magnitudes are good proxies of the rest-frame $M_{\nuv}$ luminosity. They show that the \oii LF can be reproduced well using the $\log L_{\oiirm} - M_{\mathrm{u}}(M_{\mathrm{g}})$ relation especially for the {\tt\string SAGE} model \citep{2016ApJS..222...22C}. It is worth mentioning that the slope $-0.373 (-0.342)$ and intercept $34.01 (34.29)$ of the linear relationship $\log L_{\oiirm} - M_{\mathrm{u}}(M_{\mathrm{g}})$ derived from the {\tt\string SAGE} model are quite close to our observational results. Therefore, our observed $L_{\oiirm} - M_{\nuv}$ relation can be used to calibrate the SAM of galaxy formation.

In APPENDIX \ref{sec:skew_shape}, we check the effect of $SNR$ threshold and cosmic variance on the $p^{\mathrm{int}}\left(L_{\oiirm}|M_{\nuv}\right)$. As shown in Figure \ref{fig:plot_three_field}, the intrinsic $L_{\oiirm}$ conditional distributions derived from different $SNR$ thresholds or different sky fields all present similar skewed shapes. In addition, although we have selected a complete sample and corrected the measurement effects as much as possible, the conditional distribution function $p^{\mathrm{int}}\left(L_{\oiirm}|M_{\nuv}\right)$ at very faint $M_{\nuv}$ cannot be well constrained if those galaxies are undetected. The future PFS survey with high quality spectra will provide opportunities to further test our result on faint galaxies.

\begin{figure*}
	\centering
	\includegraphics[scale=0.6]{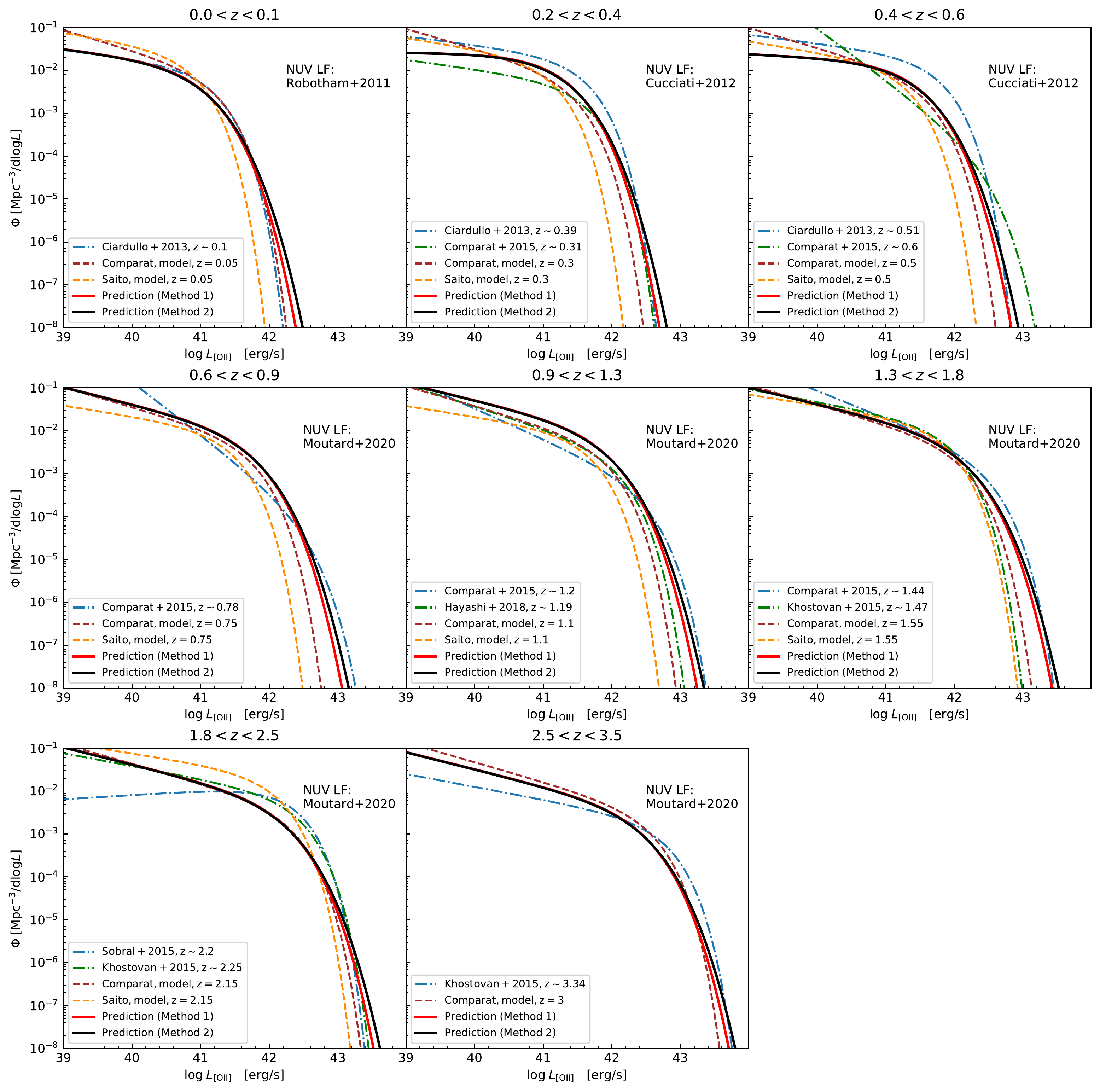}
	\caption{The comparison of our predicted \oii luminosity function (LF) with the observational results from the literature at different redshift. The red and black solid lines show our predicted \oii LFs, derived by convolving the NUV LFs measured by \cite{2011MNRAS.413.2570R}, \cite{2012A&A...539A..31C} and \cite{2020MNRAS.tmp..668M} in different redshift intervals with our best-fit model $p^{\mathrm{int}}\left(L_{\oiirm}|M_{\nuv}\right)$ obtained by method 1 and method 2, respectively. The dash-dotted lines with different colors represent the observed LFs calculated with the the best-fit Schechter \citep{1976ApJ...203..297S} parameters in some recent measurements \citep{2013ApJ...769...83C, 2015A&A...575A..40C, 2015MNRAS.451.2303S, 2015MNRAS.452.3948K, 2018PASJ...70S..17H}. Additionally, using two redshift dependent Schechter LF models proposed by \cite{2016MNRAS.461.1076C} and \cite{2020MNRAS.494..199S}, we calculate \oii LFs at the mean redshift of each $z$ interval, and plot them as brown and orange dashed lines in each panel respectively. \label{fig:LF}}
\end{figure*}  

\begin{figure}
	\centering
	\includegraphics[scale=0.7]{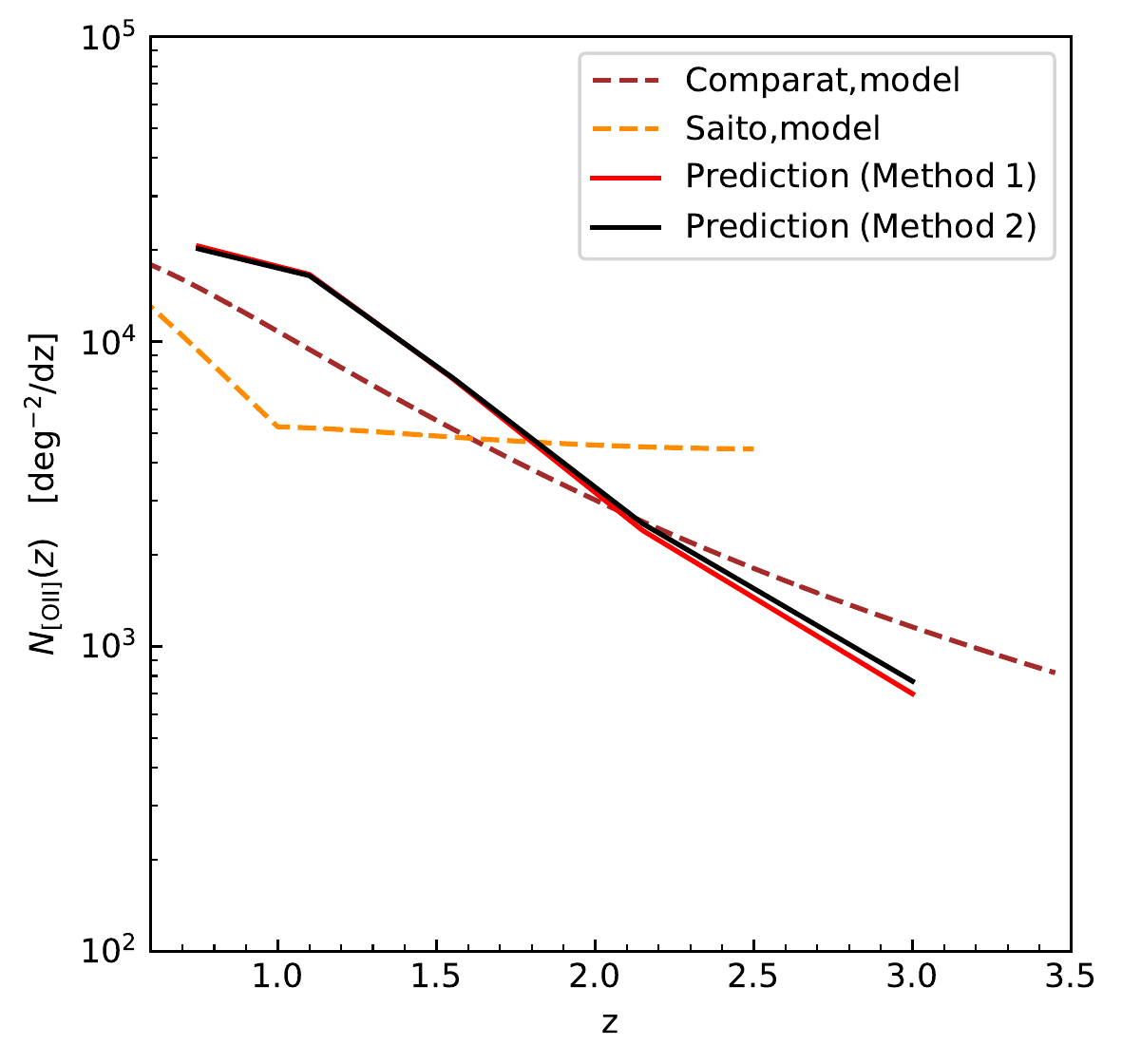}
	\caption{The prediction for the number counts of galaxies with \oii flux greater than $6.3 \times 10^{-17}\mathrm{erg s^{-1} cm^{-2}}$. Two solid lines correspond to our predictions based on our two methods, which nearly overlap each other. The brown and orange dashed lines represent the $N_{\oiirm}$ calculated based on the \oii LF models developed by \cite{2016MNRAS.461.1076C} and \cite{2020MNRAS.494..199S} respectively. The model of \cite{2020MNRAS.494..199S} was calibrated up to redshift 2.5 only by the authors.\label{fig:count}}
\end{figure}  

\section{Prediction for the luminosity function and count of the \oii emitters} \label{sec:results}

\subsection{Prediction of the \oii luminosity function} \label{subsec:predict_LF}
Convolving the intrinsic PDF $p^{\mathrm{int}}\left(L_{\oiirm}|M_{\nuv}\right)$ with NUV luminosity function $\Phi\left(M_{\nuv}\right)$, we can proceed to predict the \oii LF by
\begin{eqnarray}
\Phi\left(L_{\oiirm}\right) = \int p^{\mathrm{int}}\left(L_{\oiirm}|M_{\nuv}\right) \Phi\left(M_{\nuv}\right)\,\mathrm{d}M_{\nuv}. \nonumber \\
\end{eqnarray}
Here, we adopt three $\Phi\left(M_{\nuv}\right)$ from the literature. For the high redshift range ($z>0.6$), we choose the NUV LF recently measured by \cite{2020MNRAS.tmp..668M} based on two state-of-the-art photometric survey Canada-France-Hawaii Telescope
Large Area U-band Deep Survey \citep[CLAUDS,][]{2019MNRAS.489.5202S} and HyperSuprime-Cam Subaru Strategic Program \citep[HSC-SSP,][]{2018PASJ...70S...4A} as well as the UV photometry from the
GALEX satellite \citep{2005ApJ...619L...1M}. Although \cite{2020MNRAS.tmp..668M} has measured $\Phi\left(M_{\nuv}\right)$ at eight redshift bins from $z=0.05$ to $z=3.5$ and provides their best-fit parameters of the classical Schechter function \citep{1976ApJ...203..297S}, the relatively small sky area (18.29 $\mathrm{deg^{2}}$) and the uncertainty of photometric redshift may affect the NUV LF measurement especially in the low redshift range (because the photo-z uncertainty is proportional to $1+z$). Therefore, we also adopt other NUV LFs measured with precise spectroscopic redshift data in the lower redshift bins. One is that measured by \cite{2011MNRAS.413.2570R} at the local universe $0<z<0.1$ using the data from SDSS DR7 \citep{2009ApJS..182..543A} and GALEX MIS \citep{2007ApJS..173..682M}. Another is that given by \cite{2012A&A...539A..31C} based on VVDS survey \citep{2005A&A...439..845L,2013A&A...559A..14L}, for which we use the best-fit Schechter parameters in redshift bin $0.2<z<0.4$ and $0.4<z<0.6$. 

The red and black solid lines in Figure \ref{fig:LF} are our predicted \oii LFs from method 1 and method 2 respectively. By comparison, the method 2 predicts more luminous \oii emitters, albeit the difference is very small. For the purpose of comparing our prediction with the observation, the \oii LFs calculated with the best-fit Schechter \citep{1976ApJ...203..297S} parameters obtained by recent observational studies \citep{2013ApJ...769...83C, 2015A&A...575A..40C, 2015MNRAS.451.2303S, 2015MNRAS.452.3948K, 2018PASJ...70S..17H} are presented in Figure \ref{fig:LF}. In particular, \cite{2016MNRAS.461.1076C} and \cite{2020MNRAS.494..199S} have modeled the \oii LF as a function of redshift, thus enabling us to calculate the \oii LF at the mean redshift in each interval. As displayed in Figure \ref{fig:LF}, for the lowest redshift bin $0<z<0.1$, the predicted \oii LF tends to slightly over-predict the number density of very luminous \oii emitters, while for the redshift intervals $0.6<z<0.9$ and $0.9<z<1.3$ our prediction shows a slightly higher number density of galaxies for $\log L_{\oiirm}<42$. Nevertheless, given the large uncertainties and (or) variations of the current observed \oii LFs, our predictions overall agree rather well with the observations in the entire redshift range $z<3.5$. Specifically, even for the highest redshift bin $2.5<z<3.5$, the predicted \oii LF is still close to the observation, which further supports that the intrinsic distribution of $L_{\oiirm}$ at given $M_{\nuv}$ is likely to be universal.

\subsection{Prediction of \oii number counts} \label{subsec:predict_number}
Furthermore, using our predicted \oii LFs at $z<3.5$, we calculate the number counts per $\mathrm{deg^{2}}$ per redshift for \oii emitters. We adopt the same flux limit $F_{\mathrm{lim}} = 6.3 \times 10^{-17}\mathrm{erg s^{-1} cm^{-2}}$ as \cite{2020MNRAS.494..199S}, where the $F_{\mathrm{lim}}$ is about six times the averaged noise expected for the PFS survey. Our two methods turn out nearly identical predictions. For comparison, we also compute the $N_{\oiirm}(z)$ based on two empirical \oii LF models proposed by \cite{2016MNRAS.461.1076C} and \cite{2020MNRAS.494..199S}. As displayed in Figure \ref{fig:count}, our predicted $N_{\oiirm} - z$ is quite close to the model of \cite{2016MNRAS.461.1076C}, though our prediction has a slightly steeper slope. In the redshift range $0.6<z<1.6$, our model predicts more \oii emitters, while our model prediction drops more rapidly than the other two models for $z>2$. Especially for redshift $z=2.5$ which is of high interest to the PFS survey, our prediction is a factor 5 lower than the model of \cite{2020MNRAS.494..199S}, and is about 30\% lower than the model of \cite{2016MNRAS.461.1076C}. 

\section{Summary} \label{sec:summary}
In this study, we construct the intrinsic connection between the \oii emission line luminosity $L_{\oiirm}$ and the rest-frame near-ultraviolet band absolute magnitude $M_{\nuv}$ based on a large sample of galaxies from the VIPERS survey. We summarize our main results as follows:
\begin{enumerate}
	\item By analyzing the calibrated spectra, we have measured the \oii flux for 54166 galaxies in the redshift range of $0.6\leq z <1.45$. Combining the eight-band photometric data, we also perform the SED template fitting to obtain the rest-frame NUV absolute magnitude $M_{\nuv}$ for each galaxy.     
	\item We propose an intrinsic conditional PDF model $p^{\mathrm{int}}\left(L_{\oiirm}|M_{\nuv}\right)$ to describe the probability distribution of $L_{\oiirm}$ at a given $M_{\nuv}$. This model is constructed by a linear relationship of $\log L_{\oiirm} - M_{\nuv}$ with a skew-lognormal distribution of $L_{\oiirm}$, and can be characterized by three parameters. We develop two different methods to carefully correct for the incompleteness and measurement uncertainty of $L_{\oiirm}$. Having accounted for these observational effects in our likelihood analysis, we derive the best-fit intrinsic model parameters through an MCMC approach. 
	\item Comparing the best-fit model with the observed data at different $z$, we find that the mean linear relationship $\log L_{\oiirm} - M_{\nuv}$ is almost redshift independent. The constant slope $k=-0.4$ indicates that the luminosity of \oii is proportional to that of NUV. To further investigate the probability distribution of $L_{\oiirm}$ at given $M_{\nuv}$, we divide galaxies into various $M_{\nuv}$ and redshift bins, and compare the observed distribution of $L_{\oiirm}$ with our predicted $p^{\mathrm{obs}}\left(L_{\oiirm}|M_{\nuv}\right)$ from the best-fit model. The comparison demonstrates that the both methods can yield the universal conditional PDF of $L_{\oiirm}$, which depends on neither NUV luminosity nor redshift. This $L_{\oiirm} - M_{\nuv}$ relation can complement the recent research of \cite{2020MNRAS.497.5432F} based on simulation, and provides a feasible approach to calibrate the SAM models.
	\item Convolving the $L_{\oiirm}$ conditional PDF with the NUV LFs adopted from the literature, we have predicted the \oii LFs at eight redshift bins spanning from $z=0$ to 3.5. Our predicted \oii LFs are broadly consistent with the observational results from previous studies, though the published $L_{\oiirm}$ LFs often have significant variations. It further supports that the conditional PDF of $L_{\oiirm}$ is universal. We also have estimated the number counts $N_{\oiirm}(z)$ of \oii emitters for the forthcoming PFS survey at the flux detection limit of $6.3 \times 10^{-17}\mathrm{erg s^{-1} cm^{-2}}$. The predicted $N_{\oiirm}(z)$ is close to that calculated by the model of \cite{2016MNRAS.461.1076C}. At $z=2.5$ which is of high interest to the PFS survey, our predicted number count is 5 times lower than the model of \cite{2020MNRAS.494..199S}.
\end{enumerate}

In conclusion, the universal conditional PDF of $L_{\oiirm}$ can be used to efficiently pre-select candidates for bright \oii emiiters, and it thus will play a significant role in optimizing the source selection strategy for future galaxy redshift surveys. Moreover, this universal $L_{\oiirm}-M_{\nuv}$ distribution function directly constructs the intrinsic relationship between the ultraviolet radiation and the \oii line emission for star-forming galaxies. It will also help us to understand the formation mechanism of the \oii emission in galaxies.

\acknowledgments
H.Y.G is truly grateful to Huanian Zhang for his kind help in measuring the flux of \oii emission lines. H.Y.G also thanks Jiaxin Han for useful discussions about Bayesian analysis. The work is supported by NSFC (11533006, 11621303, 11890691) and by 111 project No. B20019. We gratefully acknowledge the support of the Key Laboratory for Particle Physics, Astrophysics and Cosmology, Ministry of Education.

This paper uses data from the VIMOS Public Extragalactic Redshift Survey (VIPERS). VIPERS has been
performed using the ESO Very Large Telescope, under the "Large Programme" 182.A-0886. The participating institutions and funding agencies are listed at
http://vipers.inaf.it. 
Based on observations collected at the European Southern Observatory, Cerro Paranal, Chile, using the Very Large Telescope under
programs 182.A-0886 and partly 070.A-9007. Also based on observations obtained with MegaPrime/MegaCam, a joint project of CFHT
and CEA/DAPNIA, at the Canada-France-Hawaii Telescope (CFHT),
which is operated by the National Research Council (NRC) of Canada,
the Institut National des Sciences de l’Univers of the Centre National
de la Recherche Scientifique (CNRS) of France, and the University of
Hawaii. This work is based in part on data products produced at TERAPIX and the Canadian Astronomy Data Centre as part of the CanadaFrance-Hawaii Telescope Legacy Survey, a collaborative project of
NRC and CNRS. This research uses data from the VIMOS VLT Deep Survey, obtained from the VVDS database operated by Cesam, Laboratoire d'Astrophysique de Marseille, France.

\software{Numpy \citep{5725236}, Scipy \citep{4160250}, Matplotlib \citep{4160265}, Astropy \citep{2013A&A...558A..33A},  emcee \citep{2013PASP..125..306F}}

\appendix
\section{The checking of \oii flux and statistics of the galaxy samples}
\label{sec:check}

\begin{figure}
	\centering
	\includegraphics[scale=0.6]{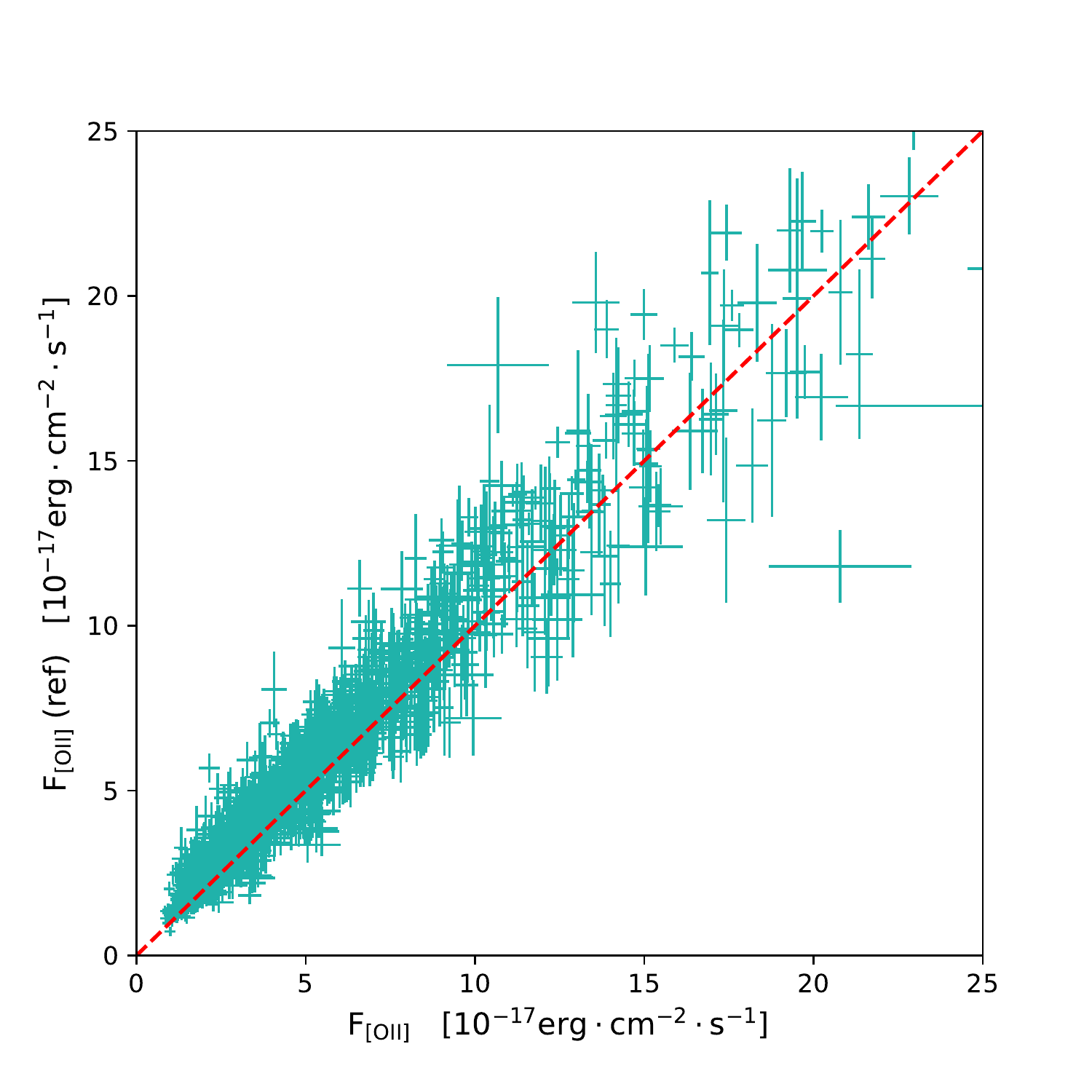}
	\caption{A check of our \oii flux measurement method. The $F_{\oiirm}$ is the \oii flux measured with our method for galaxies in the VVDS-22h wide field (F22) of the VIMOS VLT Deep Survey \citep[VVDS,][]{2005A&A...439..845L,2013A&A...559A..14L}, compared with the $F_{\oiirm} \; (\mathrm{ref})$ of the previous measurement from \cite{2009A&A...495...53L}. The measurements with $SNR>5$ are displayed together as green points with horizontal and vertical error bars. The red dashed line represents that $F_{\oiirm}\;(\mathrm{ref})=F_{\oiirm}$ .\label{fig:flux_vvds_compare}}
\end{figure}

\cite{2009A&A...495...53L} have released their measured \oii flux in the VVDS-22h wide field (F22) of the VIMOS VLT Deep Survey \citep[VVDS,][]{2005A&A...439..845L,2013A&A...559A..14L}, which enables us to check our method for the \oii measurement. We apply our method to the same galaxies in the VVDS-22h wide field (F22) to measure their \oii flux. A comparison between the two measurements is presented in Figure \ref{fig:flux_vvds_compare}, which shows a very good agreement. We note a small systematic difference (less than $10\%$) at small flux which may be attributed to the difference of the two studies in subtracting the continuum, as \cite{2009A&A...495...53L} used a combination of stellar population templates to model the stellar component of the spectra but we use a sixth-order polynomial to describe the continuum around the \oii line.

We apply this method to the VIPERS galaxies with redshift flag $\geq 2$ in the range of $0.6\leq z<1.45$. For those weak or noisy spectral lines, the fitting may fail, and/or the covariance matrix of the parameters may not be calculated correctly. Here, we define a line fitting as being successful if following three conditions are satisfied: (a) $F_{\oiirm}>0$, (b) the diagonal elements of the covariance matrix are positive, and (c) the uncertainty of line width $\sigma_{W_{\oiirm}}<100$\AA (the fitted line widths $W_{\oiirm}$ are normally between 3-10\AA, so we remove those with $\sigma_{W_{\oiirm}}\geq 100$\AA ~that are obviously outliers). As the result, 45139 out of the total 54166 galaxies have been successfully fitted.

\begin{figure*}
	\centering
	\includegraphics[scale=0.8]{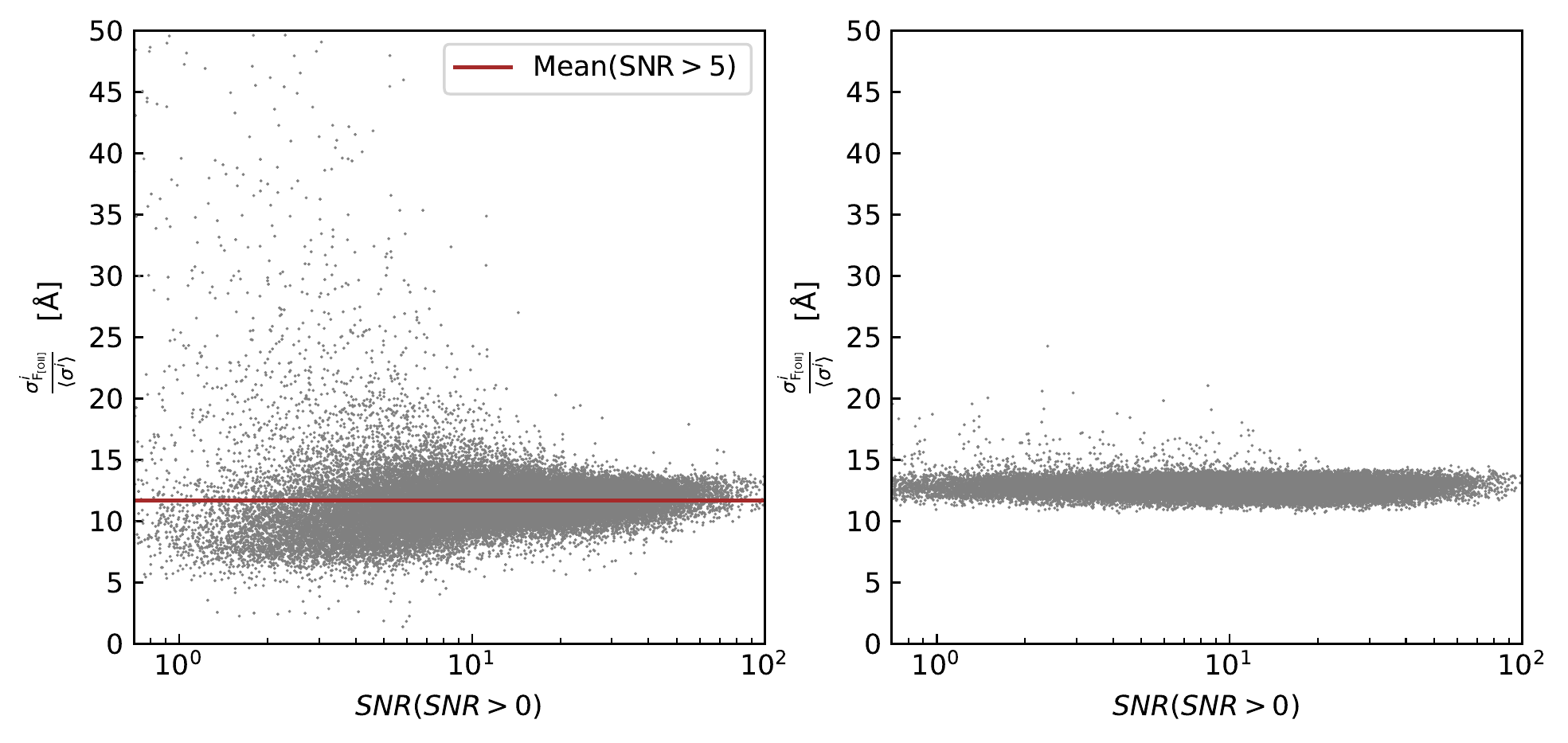}
	\caption{The ratio of the flux uncertainty $\sigma^i_{F_{\oiirm}}$ to the average background noise $\left \langle \sigma^{i} \right \rangle$ in the range of $3727 \pm 20 \rm{\AA}$ as a function of $SNR$. Each gray point corresponds to a galaxy in our VIPERS sample. The $\sigma^i_{F_{\oiirm}}$ in the left panel and right panel is estimated from the covariance matrix and the error propagation of noise spectrum, respectively. The solid brown line in the left panel shows the mean ratio $\left \langle\sigma^i_{F_{\oiirm}}/\left \langle \sigma^{i} \right \rangle\right \rangle = 11.653 \rm{\AA}$ for the galaxies with $SNR>5$. \label{fig:noise_distribution}}
\end{figure*}

Nonetheless, it should be emphasized that in our method 2 (see Equation \ref{equ:Cobs} and \ref{equ:ln2}), we use all galaxies, not only those fitted successfully ($SNR>0$), to ensure that our sample is complete. Therefore, for the purpose of determining the $SNR$ threshold (e.g. $L^{i}_{\oiirm,\mathrm{thr}}=5\sigma^i _{L_{\oiirm}}$), we also need to estimate the measurement uncertainty for the lines that failed to be fitted. In order to estimate their $\sigma_{F_{\oiirm}}$, we calculate the ratio of $\sigma^i_{F_{\oiirm}}$ to $\left \langle \sigma^i \right \rangle$ for each galaxies with $SNR>0$, here the $\sigma^i_{F_{\oiirm}}$ is derived from the covariance matrix and the $\left \langle \sigma^i \right \rangle$ is the average of noise spectrum in the fitting range ($3727 \pm 20 \rm{\AA}$) for the {\it i}-th galaxy. As displayed in the left panel of Figure \ref{fig:noise_distribution}, $\sigma^i_{F_{\oiirm}}$ is roughly proportional to $\left \langle \sigma^i \right \rangle$ for the galaxies with high $SNR$. The mean ratio $\left \langle\sigma^i_{F_{\oiirm}}/\left \langle \sigma^{i} \right \rangle\right \rangle \left(SNR>5\right)= 11.653 \rm{\AA}$  is plotted as a solid brown line in Figure \ref{fig:noise_distribution}. Hence it enables us to approximately estimate the $\sigma^i_{F_{\oiirm}}$ through multiplying this mean ratio by the average background noise ($\sigma^i_{F_{\oiirm}} \simeq 11.653\rm{\AA}\times \left \langle \sigma^{i} \right \rangle$) for those galaxies that failed being fitted. As a note, one may also estimate the $\sigma^i_{F_{\oiirm}}$ from the error propagation of noise spectrum (the root-sum-square of noise spectrum in the range of $3727 \pm 20 \rm{\AA}$ times the spacing of pixels $\Delta \lambda$), and we show it in the right panel of Figure \ref{fig:noise_distribution}. The two estimates are similar for the \oii with high $SNR$, but the errors from the covariance matrix are more dispersive at low $SNR$, reflecting the additional errors associated with the measurement of the \oii line. We used these two types of errors in our statistical study of the relation between \oii and NUV luminosities, and found that the result is insensitive to which type of error used.  In this paper, we adopt the $\sigma^i_{F_{\oiirm}}$ estimated from the covariance matrix.

\begin{figure*}
	\centering
	\includegraphics[scale=0.6]{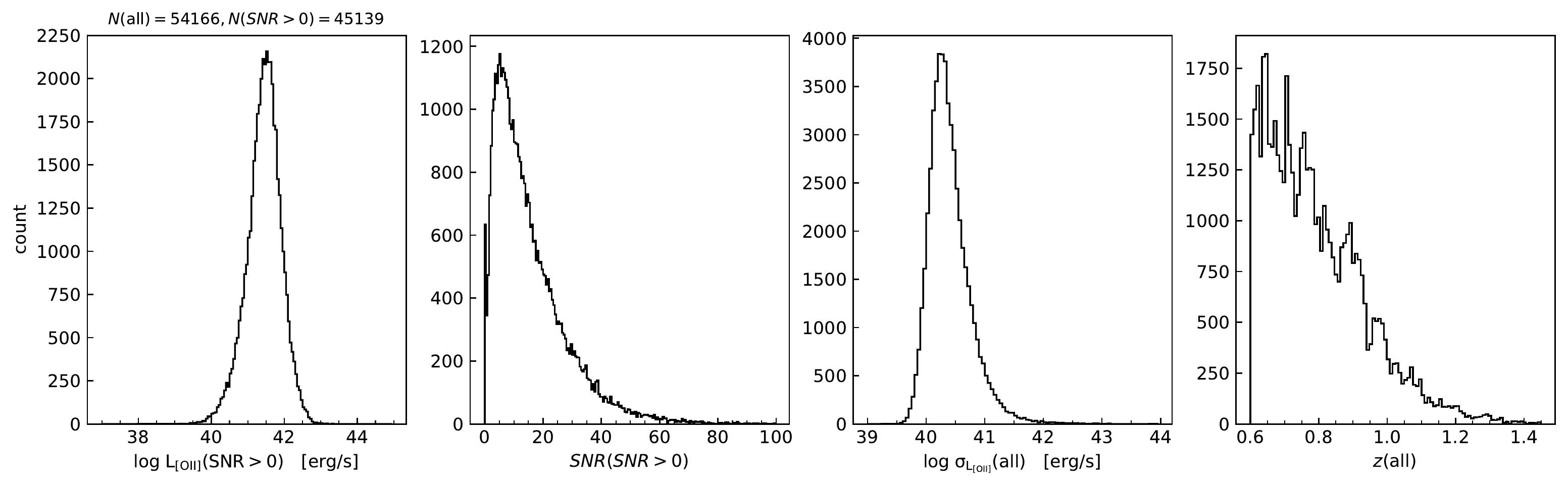}
	\caption{The distributions of \oii luminosity $L_{\oiirm}$, $SNR$, uncertainty $\sigma_{L_{\oiirm}}$ and redshift of our galaxy sample.\label{fig:data_distribution}}
\end{figure*}

Finally, in Figure \ref{fig:data_distribution}, we present the distributions of \oii luminosity, $SNR$, uncertainty and redshift of galaxies in our sample.

\section{Test the effect of signal-to-noise ratio and cosmic variance on the skewed shape of $L_{\oiirm}$ conditional distribution} \label{sec:skew_shape}

\begin{figure}
	\centering
	\includegraphics[scale=0.8]{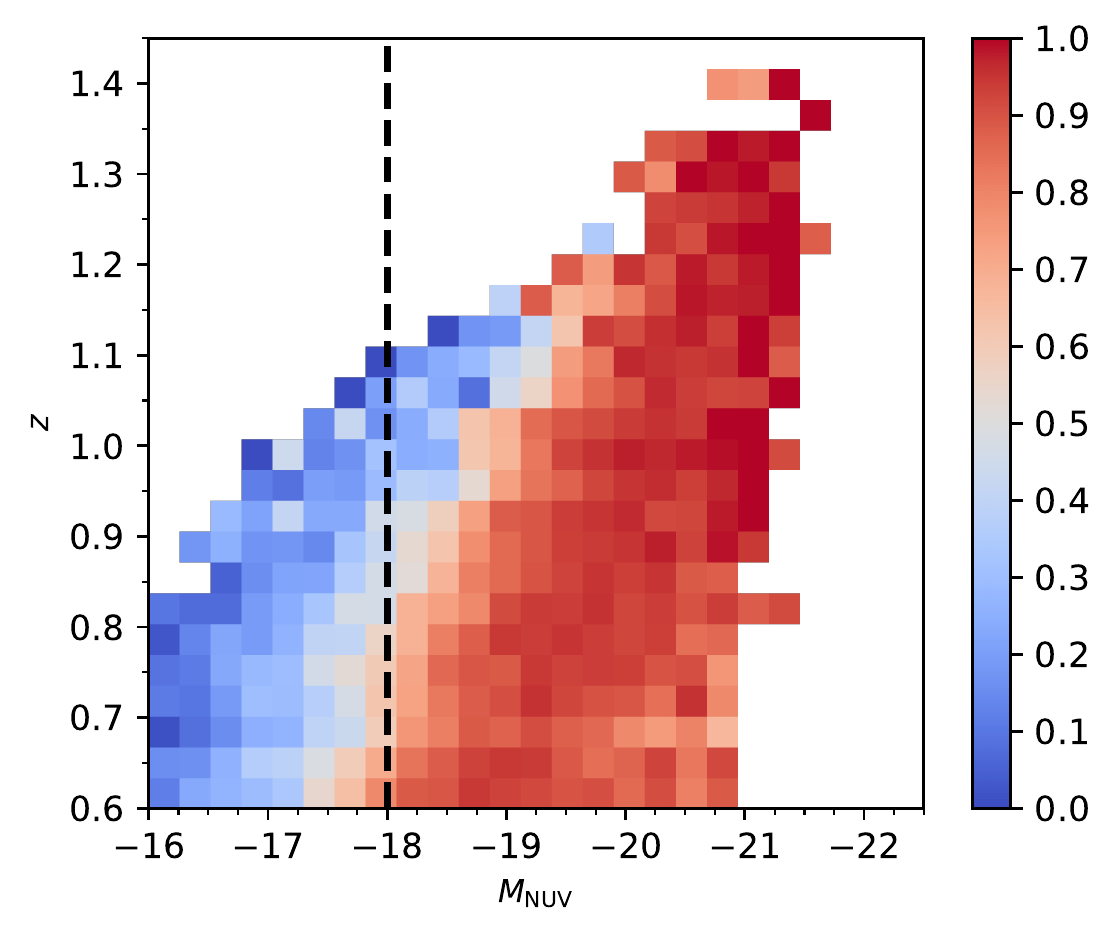}
	\caption{Similar to Figure \ref{fig:MeasureSuccessRate2d}, the panel shows $MSR^{SNR>3} = N^w(SNR>3)/N^w$.\label{fig:Measure_success_rate_2d_SNR3}}
\end{figure}

In this appendix, we investigate two factors that may impact the intrinsic shape of the $L_{\oiirm}$ conditional probability distribution function (PDF for short). One is the specific values (1 and 5) set for  the $SNR$ threshold in  method 1 and method 2 respectively (see also Section \ref{sec:methodology}). We should carefully analyze the influence of this factor on the shape of the PDF at the  faint end. Considering that some spurious \oii emitters with $SNR<5$ could affect the purity of our samples, we adopt $SNR=3$ as a new threshold in method 2. The $MSR$ of our samples is shown in Figure \ref{fig:Measure_success_rate_2d_SNR3}.

The other factor is the effect of a finite survey volume or sample size, usually called as the cosmic variance. The cosmic variance could more likely affect the shape of \oii PDF at the bright end where the number of luminous galaxies is small. Therefore, we divide our galaxy sample into three fields: W1(RA$\leq34.5\;$deg), W1(RA$>34.5\;$deg) and W4, and repeat the method 2 to derive the intrinsic \oii PDF in each field. The results are displayed in in Table \ref{tab:result_test}.

\begin{figure}
	\centering
	\includegraphics[scale=0.6]{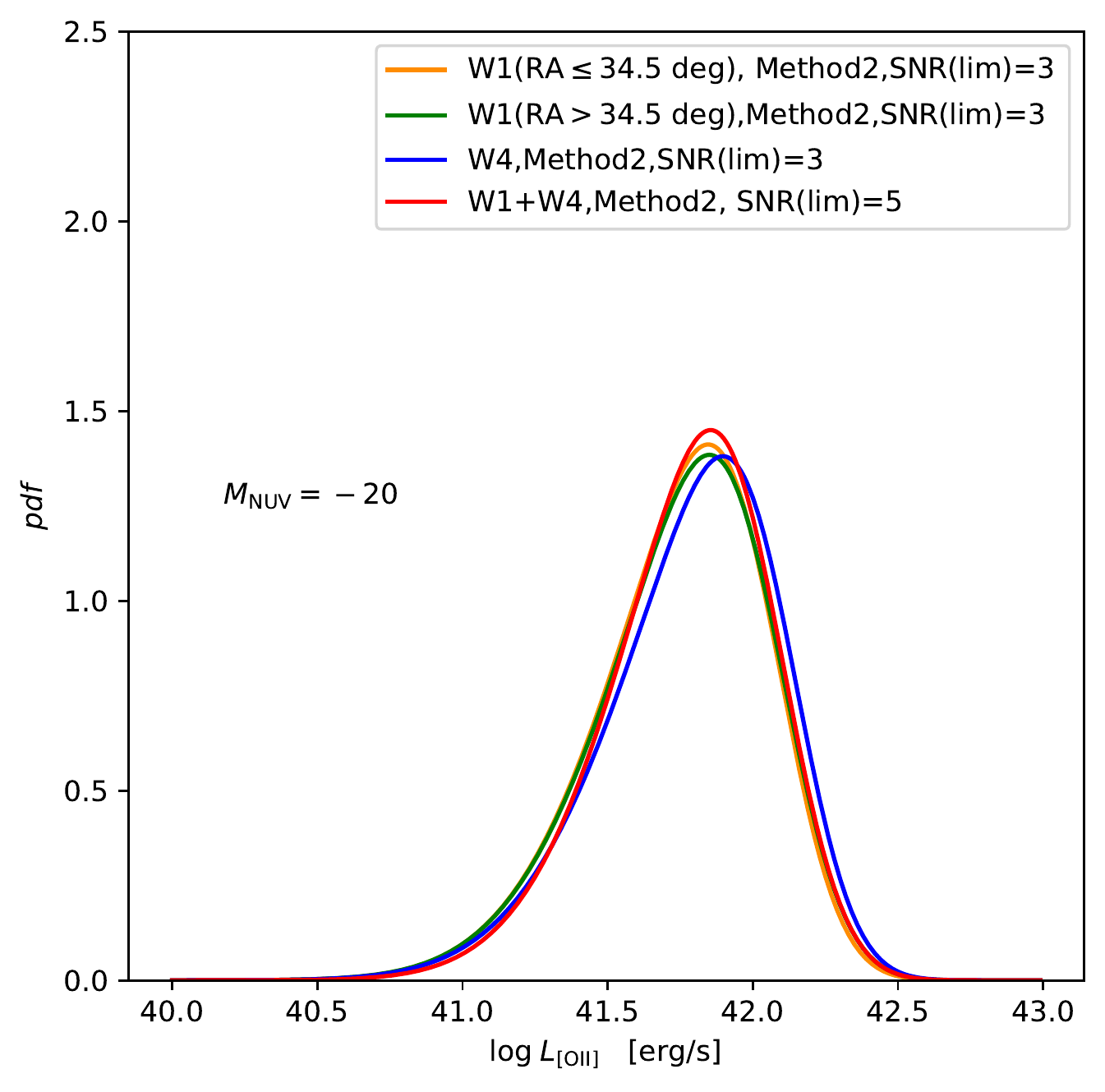}
	\caption{The intrinsic PDF $p^{\mathrm{int}}\left(L_{\oiirm}|M_{\nuv}\right)$ calculated with the best-fitting parameters at $M_{\nuv}=-20$. The orange, green and blue lines corresponding to the three rows of Table \ref{tab:result_test} represent the PDF derived at $SNR>3$ in three different fields. For comparison, the PDF derived at $SNR>5$ using all galaxies (the last row of Table \ref{tab:result}) is plotted as a red line. \label{fig:plot_three_field}}
\end{figure}

We plot the three intrinsic PDF $p^{\mathrm{int}}\left(L_{\oiirm}|M_{\nuv}\right)$ calculated at $M_{\nuv}=-20$ in Figure \ref{fig:plot_three_field}. By comparison, the intrinsic \oii PDF in all three fields exhibit similar skew-normal ($\alpha<0$) shape, while the expectation value $\log L^{\mathrm{exp}}_{\oiirm}$ in W4 is about 0.05 dex (see also Table \ref{tab:result_test}) brighter than that in W1. This means that cosmic variance hardly affects the shape of the \oii PDF, but has a slight effect on the intercept of the $\log L_{\oiirm} - M_{\nuv}$ relation. On the other hand, to explore the influence of the $SNR$ threshold, we also plot the intrinsic \oii PDF derived from the threshold of $SNR=5$ (the last row of Table \ref{tab:result}) as a red line in Figure \ref{fig:plot_three_field}. Although the PDF derived from $SNR>5$ is slightly more concentrated than that derived from $SNR>3$, they both show a skew-normal shape extending to the faint side.

We have also checked the sensitivity of our result to the $SNR$ threshold in method 1. We repeated the calculation by setting $SNR>0$, and found the result has little changed.

Therefore, our results are robust both to the reasonable values set for the $SNR$ thresholds and to the cosmic variance as the sample is sufficiently large.

\begin{deluxetable*}{ccccccccc}
	\tablenum{2}
	\tablecaption{Similar to Table \ref{tab:result}. The best-fitting model parameters with their $1\sigma$ uncertainties are derived from galaxies with $SNR$ threshold equal to 3 in three survey fields. \label{tab:result_test}}
	\tablehead{
		\colhead{Method} & \colhead{Redshift Range} & \colhead{Field} &\colhead{$N_{\mathrm{galaxy}}$} & \colhead{$N^{SNR>3}_{\mathrm{galaxy}}$} &
		\colhead{$\alpha$} & \colhead{$\omega$} & \colhead{$k$} & \colhead{$b$}
	}
	\startdata
	{        }&$0.6\leq z<1.45$& W1(RA$\leq34.5\;$deg)& 14723 &12927& $-2.4109^{+0.0363}_{-0.0363}$&$0.4419^{+0.0028}_{-0.0027}$&$-0.4$&$34.0711^{+0.0023}_{-0.0024}$\\
	{Method 2}&$0.6\leq z<1.45$& W1(RA$>34.5\;$deg)&14099 &12418&$-2.3235^{+0.0361}_{-0.0367}$&$0.4463^{+0.0029}_{-0.0028}$&$-0.4$&$34.0797^{+0.0026}_{-0.0025}$\\
	{        }&$0.6\leq z<1.45$&W4&13570&11691&$-2.5648^{+0.0379}_{-0.0385}$&$0.4590^{+0.0029}_{-0.0029}$&$-0.4$&$34.1263^{+0.0023}_{-0.0024}$\\
	\enddata
\end{deluxetable*}

\bibliography{oii}{}
\bibliographystyle{aasjournal}

\end{document}